\documentclass[aps,prd,twocolumn,superscriptaddress]{revtex4}
\usepackage{epsfig,epsf}
\usepackage{amsmath}
\usepackage{amsthm}
\usepackage{amsfonts}
\usepackage{amssymb}
\usepackage{dsfont}
\usepackage{multirow}
\usepackage{appendix}
\usepackage{slashed}
\usepackage[active]{srcltx}
\usepackage{psfrag}

\begin{document}

\title{Decay widths of the excited $\Omega_b$ baryons}
\date{\today}
\author{S.~S.~Agaev}
\affiliation{Institute for Physical Problems, Baku State University, Az--1148 Baku,
Azerbaijan}
\author{K.~Azizi}
\affiliation{School of Physics, Institute for Research in Fundamental Sciences (IPM),\\
P.O.Box 19395-5531, Tehran, Iran}
\affiliation{Department of Physics, Do\v{g}u\c{s} University, Acibadem-Kadik\"{o}y, 34722
Istanbul, Turkey}
\author{H.~Sundu}
\affiliation{Department of Physics, Kocaeli University, 41380 Izmit, Turkey}

\begin{abstract}
The LHCb Collaboration recently observed five narrow $\Omega_c^{0}$
resonances, and measured their masses and widths through the decays $%
\Omega_c^{0} \to \Xi_c^{+}K^{-}$. Motivated by this discovery, and also by
the fact that the ground-state bottom baryon $\Omega_b^{-}$ with spin-1/2
was already found experimentally, we perform theoretical investigation of
the spin-1/2 and spin-3/2, $\Omega_b$, baryons by calculating decay width of
their first orbitally and radially excited states to $\Xi_b^{0} K^{-}$. For
this purpose, we employ QCD sum rule method on the light-cone by including
into analysis the $K$ meson distribution amplitudes up to twist-4. Obtained
analytical expressions are utilized to extract parameters of these decay
processes which may be useful for forthcoming experimental studies of bottom
baryons.
\end{abstract}

\maketitle


\section{Introduction}

\label{sec:Int} 

Recent discovery of the five narrow $\Omega_c^{0} $ states \cite%
{Aaij:2017nav}, and observation of the double charmed baryon $\Xi_{cc}^{++}$
by the LHCb Collaboration \cite{Aaij:2017ueg} opened new page in the
experimental physics of heavy flavored baryons. They also stimulated new and
more detailed theoretical studies of baryons containing one or two heavy
quarks which has become one of interesting areas of high energy physics. In
fact, variety of interpretations were proposed in Refs.\ \cite%
{Agaev:2017jyt,Agaev:2017lip,Chen:2017sci,Karliner:2017kfm,Wang:2017vnc,Padmanath:2017lng, Yang:2017rpg,Huang:2017dwn,Kim:2017jpx,Wang:2017hej,Cheng:2017ove,Wang:2017zjw,Aliev:2017led}
to understand the nature of the observed $\Omega_c^{0}$ resonances: They
were considered as $P$-wave charmed baryons $\Omega_c^{0}$ of different
spins, as the orbitally and radially excited states of spin-1/2 and spin-3/2
particles $\Omega_c^{0}$ and $\Omega_c^{\star 0}$, or even as pentaquark
candidates. Additional information on suggested explanations and references
to corresponding works can be found in Ref.\ \cite{Agaev:2017lip}.

As is seen experimental investigations of the charmed $\Omega_c$ or double
charmed baryons have achieved remarkable successes, whereas the bottom
baryons $\Omega_b$ suffer from deficiency of experimental data. Indeed, in
the class of $\Omega_b^{-}$ baryons the data are restricted by the mass of
the spin-1/2 baryon $\Omega _{b}^{-}$ (see, Ref.\ \cite{Olive:2016xmw})
\begin{equation}
m=6071 \pm 40\,\mathrm{\,MeV}.  \label{eq:Data2}
\end{equation}

On contrary, theoretical studies of the bottom baryons encompass variety
of models and methods. The spectra of the ground and excited states of the
heavy flavored baryons were studied in the context of the QCD sum rule
method \cite%
{Bagan:1991sc,Bagan:1992tp,Huang:2000tn,Wang:2002ts,Wang:2007sqa,Wang:2008hz,Wang:2009cr, Wang:2010vn,Wang:2011, Mao:2015gya,Chen:2016phw,Aliev:2016jnp}%
, different relativistic and non-relativistic quark models \cite%
{Capstick:1986bm,Ebert:2007nw,Ebert:2011kk,Garcilazo:2007eh,
Valcarce:2008dr,Roberts:2007ni,Vijande:2012mk,Yoshida:2015tia}. The magnetic
moments, radiative decays, strong couplings and radiative transitions of the
heavy flavored baryons were subject of intensive theoretical studies, as
well \cite{Aliev:2008ay,
Aliev:2008sk,Aliev:2009jt,Aliev:2010nh,Aliev:2010ev,Aliev:2011kn,Aliev:2011ufa,Aliev:2011uf}%
. Sometimes it is difficult to classify uniquely these works basing only on
the used methods or assumptions made on the structures of baryons because most of
them combines different models and computational schemes. For example, in
the relativistic quark model baryons were considered as the
heavy-quark--light-diquark bound states \cite{Ebert:2007nw,Ebert:2011kk}. In
other papers, QCD sum rule calculations were supplied by methods of the
heavy quark effective theory \cite{Huang:2000tn,Wang:2002ts,Chen:2016phw}.

New experimental situation necessities a detailed exploration of the $\Omega
_{b} $ baryons which should embrace parameters  of the ground-state and
excited baryons, as well as their possible decay channels. As it has been
just noted mass spectra of the bottom baryons were studied in numerous
works. Recently, in the context of the different approaches these problems
were revisited in Refs.\ \cite{Agaev:2017jyt,Thakkar:2016dna}. Thus, masses
and pole residues of the ground-state and excited $\Omega _{b}=(1S,\
1/2^{+}) $, $\widetilde{\Omega }_{b}=(1P,1/2^{-})$, $\Omega _{b}^{\prime
}=(2S,1/2^{+})$ and $\Omega _{b}^{\star }=(1S,\ 3/2^{+})$, $\widetilde{%
\Omega }_{b}^{\star }=(1P,3/2^{-})$, $\Omega _{b}^{\star \prime
}=(2S,3/2^{+})$ \ baryons (hereafter, for the sake of simplicity we omit in
notations a superscript "$-$") were calculated in the framework of QCD
two-point sum rule method in Ref.\ \cite{Agaev:2017jyt}. The questions of
mass spectra of excited $\Sigma _{b}$, $\Lambda _{b}$ and $\Omega _{b}$
baryons in the context of the hypercentral constituent quark model were
addressed in Ref.\ \cite{Thakkar:2016dna}, where authors analyzed also
semi-electronic decays of the $\Omega _{b}$ and $\Sigma _{b}$ baryons. The
properties of the $D$-wave heavy baryons were considered in Ref.\ \cite%
{Mao:2017wbz}.

In the present work we extend our previous investigation \cite{Agaev:2017jyt}
and calculate the width of strong decays of $\Omega _{b}$ and $\Omega
_{b}^{\star }$ baryons to $\Xi _{b}^{0}K^{-}$. We are going to follow a
scheme applied in Ref.\ \cite{Agaev:2017lip} to study decays of the excited
spin-1/2 and spin-3/2 baryons $\Omega _{c}$ and $\Omega _{c}^{\star }$ . It
turns out that, as in the case of $\Omega _{c}$ and $\Omega _{c}^{\star }$,
only decays of orbitally and radially excited baryons $\widetilde{%
\Omega }_{b}$, $\Omega _{b}^{\prime }$ and $\widetilde{\Omega }_{b}^{\star }$%
, $\Omega _{b}^{\star \prime }$ to $\Xi _{b}^{0}K^{-}$ are kinematically
allowed. The spectroscopic parameters of the $\Omega _{b}$ and\ $\Omega
_{b}^{\star }$ obtained in Ref.\ \cite{Agaev:2017jyt} will be applied as
input information in light-cone sum rule calculations of the strong
couplings $g_{\Omega _{b}\Xi _{b}K}$ and $g_{\Omega _{b}^{\star }\Xi _{b}K}$
which are necessary to find decay widths $\Gamma (\Omega _{b}\rightarrow \Xi
_{b}K)$ and $\Gamma (\Omega _{b}^{\star }\rightarrow \Xi _{b}K)$.

This article is structured in the following way. In Sec.\ \ref{sec:decays}
we calculate the strong couplings $g_{\Omega_b\Xi_bK}$ and $g_{\Omega_b^{\star}%
\Xi_bK}$ using of QCD light-cone sum rule method. Here we provide general
expressions for width of the corresponding decay processes. Section \ref%
{sec:NumRest} is reserved to numerical computations, where we give a
required information on parameters employed during this process, as well as
provide our predictions for the width of the decays of interest. Section \ref%
{sec:Conc} contains our concluding remarks. In Appendix we write down the Borel transformed  form  of some invariant amplitudes used in the analyses. One can find here also an information on distribution
amplitudes of $K$ meson, as well as expressions used in the continuum
subtraction.


\section{Decays of orbital and radial excitations of $\Omega_b$ and $%
\Omega_b^{\star}$ baryons to $\Xi_b^{0}K^{-}$ final state}

\label{sec:decays}
As we have noted above the masses of ground-state and excited $\Omega _{b}$
and $\Omega _{b}^{\star }$ baryons were extracted from QCD two-point sum
rules in Ref.\ \cite{Agaev:2017jyt}, where contributions of various quark,
gluon and mixed condensates up to dimension ten were taken into account. For
$J=1/2$ baryons $\Omega _{b}$, $\widetilde{\Omega }_{b}$ and $\Omega
_{b}^{\prime }$ we found (in $\mathrm{MeV}$)
\begin{equation}
m=6024\pm 183,\widetilde{m}=6336\pm 183,m^{\prime }=6487\pm 187,
\end{equation}%
whereas for $J=3/2$ baryons $\Omega _{b}^{\star }$, $\widetilde{\Omega }%
_{b}^{\star }$ and $\Omega _{b}^{\star \prime }$ we obtained
\begin{equation}
m^{\star }=6084\pm 161,\widetilde{m}^{\star }=6301\pm 193,m^{\star \prime
}=6422\pm 198.
\end{equation}%
By taking into account experimental data on masses of the particles $\Xi
_{b}^{0}$ and $K$
\begin{equation}
m_{\Xi _{b}}=5791.9\pm 0.5\,\mathrm{MeV},m_{K}=493.677\pm 0.016\ \mathrm{MeV}%
,
\end{equation}%
it is not difficult to see that only excited $\Omega _{b}$ and $\Omega
_{b}^{\star }$ baryons can decay to the final state $\Xi _{b}K$.


\subsection{$\widetilde{\Omega}_b \to \Xi_{b}^{0}K^{-}$ and $%
\Omega_b^{\prime} \to \Xi_{b}^{0}K^{-}$ decays}


We start our consideration from the strong vertices $\widetilde{\Omega }%
_{b}\Xi _{b}^{0}K^{-}$ and $\Omega _{b}^{\prime }\Xi _{b}^{0}K^{-}$, and
calculate corresponding couplings $g_{\widetilde{\Omega }_{b}\Xi _{b}K}$ and
$g_{\Omega _{b}^{\prime }\Xi _{b}K}$ , which are required to determine width
of the decays $\widetilde{\Omega }_{b}\rightarrow \Xi _{b}^{0}K^{-}$ and $%
\Omega _{b}^{\prime }\rightarrow \Xi _{b}^{0}K^{-}$. For these purposes we
explore the correlation function%
\begin{equation}
\Pi (p,q)=i\int d^{4}xe^{ipx}\langle K(q)|\mathcal{T}\{\eta _{\Xi _{b}}(x)%
\overline{\eta }(0)\}|0\rangle ,
\end{equation}%
where $\eta (x)$ and $\eta _{\Xi _{b}}(x)$ are interpolating currents for
the $\Omega _{b}$ and $\Xi _{b}^{0}$ baryons, respectively. The
interpolating current matching quantum numbers and quark content of the $%
\Omega _{b}$ baryons are given by the expression
\begin{equation}
\eta =\epsilon ^{abc}\left[ \left( b^{aT}Cs^{b}\right) \gamma
_{5}s^{c}+\beta \left( b^{aT}C\gamma _{5}s^{b}\right) s^{c}\right] ,
\label{eq:BayC1/2}
\end{equation}%
where $C$ is the charge conjugation operator. The current for spin-$1/2$
baryons $\eta (x)$ contains an arbitrary auxiliary parameter $\beta $: The
case $\beta =-1$ corresponds to the well known Ioffe current.

The baryon $\Xi _{b}^{0}$ belongs to the anti-triplet configuration of the
heavy baryons containing a single heavy quark. The relevant interpolating
current $\eta _{\Xi _{b}}$ is anti-symmetric with respect to exchange of two
light quarks, and is given by the expression
\begin{eqnarray}
\eta _{\Xi _{b}} &=&\frac{1}{\sqrt{6}}\epsilon ^{abc}\left\{ 2\left(
u^{aT}Cs^{b}\right) \gamma _{5}b^{c}+2\beta \left( u^{aT}C\gamma
_{5}s^{b}\right) b^{c}\right.  \notag \\
&&+\left( u^{aT}Cb^{b}\right) \gamma _{5}s^{c}+\beta \left( u^{aT}C\gamma
_{5}b^{b}\right) s^{c}  \notag \\
&&\left. +\left( b^{aT}Cs^{b}\right) \gamma _{5}u^{c}+\beta \left(
b^{aT}C\gamma _{5}s^{b}\right) u^{c}\right\} .  \label{eq:XiCurr}
\end{eqnarray}

As the first step we represent the correlation function $\Pi (p,q)$ using
the parameters of the involved baryons, and determine the phenomenological
side of the sum rules. To this end, we write down $\Pi (p,q)$ in the
following form:
\begin{eqnarray}
&&\Pi ^{\mathrm{Phys}}(p,q)=\frac{\langle 0|\eta _{\Xi _{b}}|\Xi
_{b}^{0}(p,s)\rangle }{p^{2}-m_{\Xi _{b}}^{2}}\langle K(q)\Xi _{b}^{0}(p,s)|%
\widetilde{\Omega }_{b}(p^{\prime },s^{\prime })\rangle  \notag \\
&&\times \frac{\langle \widetilde{\Omega }_{b}(p^{\prime },s^{\prime })|%
\overline{\eta }|0\rangle }{p^{\prime 2}-\widetilde{m}^{2}}+\frac{\langle
0|\eta _{\Xi _{b}}|\Xi _{b}^{0}(p,s)\rangle }{p^{2}-m_{\Xi _{b}}^{2}}  \notag
\\
&&\times \langle K(q)\Xi _{b}^{0}(p,s)|\Omega _{b}^{\prime }(p^{\prime
},s^{\prime })\rangle \frac{\langle \Omega _{b}^{\prime }(p^{\prime
},s^{\prime })|\overline{\eta }|0\rangle }{p^{\prime }{}^{2}-m^{\prime 2}}%
+\ldots ,  \label{eq:SRDecay}
\end{eqnarray}%
where $p^{\prime }=p+q,\ p$ and $q$ are the momenta of the $\Omega _{b},\
\Xi _{b}^{0}$ baryons and $K$ meson, respectively. The contributions of the
higher resonances and continuum states are denoted in Eq.\ (\ref{eq:SRDecay}%
) by dots.

Further simplification in Eq.\ (\ref{eq:SRDecay}) are achieved by expressing
matrix elements in terms of hadronic parameters and strong couplings. Thus,
we introduce the matrix elements of $\Omega _{b}$ and $\Xi _{b}^{0}$
baryons: for $\widetilde{\Omega }_{b}$ and $\Omega _{b}^{\prime }$ we have
\begin{eqnarray}
\langle 0|\eta |\widetilde{\Omega }_{b}(p,s)\rangle &=&\widetilde{\lambda }%
\gamma _{5}\widetilde{u}(p,s),  \notag \\
\langle 0|\eta |\Omega _{b}^{\prime }(p,s)\rangle &=&\lambda ^{\prime
}u^{\prime }(p,s),  \label{eq:MElem}
\end{eqnarray}%
where $\widetilde{\lambda }$ and $\lambda ^{\prime }$ are the pole residues
of $\widetilde{\Omega }_{b}$ and $\Omega _{b}^{\prime }$ states,
respectively. The matrix element of $\Xi _{b}^{0}$ is defined by a similar
manner
\begin{equation*}
\langle 0|\eta _{\Xi _{b}}|\Xi _{b}^{0}(p,s)\rangle =\lambda _{\Xi
_{b}}u(p,s).
\end{equation*}%
We use also the definitions for the strong couplings:
\begin{eqnarray}
&&\langle K(q)\Xi _{b}^{0}(p,s)|\widetilde{\Omega }_{b}(p^{\prime
},s^{\prime })\rangle =g_{\widetilde{\Omega }_{b}\Xi _{b}K}\overline{u}%
(p,s)u(p^{\prime },s^{\prime }),  \notag \\
&&\langle K(q)\Xi _{b}^{0}(p,s)|\Omega _{b}^{\prime }(p^{\prime },s^{\prime
})\rangle =g_{\Omega _{b}^{\prime }\Xi _{b}K}\overline{u}(p,s)\gamma
_{5}u(p^{\prime },s^{\prime }).  \notag \\
&&{}
\end{eqnarray}%
Employing these matrix elements, and carrying out the summation over $s$ and
$s^{\prime }$ in accordance with the prescription
\begin{equation}
\sum\limits_{s}u(p,s)\overline{u}(p,s)=\slashed p+m,
\end{equation}%
one can easily recast the function $\Pi ^{\mathrm{Phys}}(p,q)$ into the form:%
\begin{eqnarray}
&&\Pi ^{\mathrm{Phys}}(p,q)=-\frac{g_{\widetilde{\Omega }_{b}\Xi
_{b}K}\lambda _{\Xi _{b}}\widetilde{\lambda }}{(p^{2}-m_{\Xi
_{b}}^{2})(p^{\prime }{}^{2}-\widetilde{m}^{2})}(\slashed p+m_{\Xi _{b}})
\notag \\
&&\times \left( \slashed p+\slashed q+\widetilde{m}\right) \gamma _{5}+\frac{%
g_{\Omega _{b}^{\prime }\Xi _{b}K}\lambda _{\Xi _{b}}\lambda ^{\prime }}{%
(p^{2}-m_{\Xi _{b}}^{2})(p^{\prime }{}^{2}-m^{\prime 2})}  \notag \\
&&\times (\slashed p+m_{\Xi _{b}})\gamma _{5}\left( \slashed p+\slashed %
q+m^{\prime }\right) +\ldots .
\end{eqnarray}%
Applying the double Borel transformation on the variables $p^{2}$ and $%
p^{\prime 2}$ for $\Pi ^{\mathrm{Phys}}(p,q)$ we get
\begin{eqnarray}
&&\mathcal{B}\Pi ^{\mathrm{Phys}}(p,q)=g_{\widetilde{\Omega }_{b}\Xi
_{b}K}\lambda _{\Xi _{b}}\widetilde{\lambda }e^{-\widetilde{m}%
^{2}/M_{1}^{2}}e^{-m_{\Xi _{b}}^{2}/M_{2}^{2}}  \notag \\
&&\times \left\{ \slashed q\slashed p\gamma _{5}-m_{\Xi _{b}}\slashed %
q\gamma _{5}-\left( \widetilde{m}+m_{\Xi _{b}}\right) \slashed p\gamma
_{5}\right.  \notag \\
&&\left. +\left[ m_{K}^{2}-\widetilde{m}(\widetilde{m}+m_{\Xi _{b}})\right]
\gamma _{5}\right\} +g_{\Omega _{b}^{\prime }\Xi _{b}K}\lambda _{\Xi
_{b}}\lambda ^{\prime }  \notag \\
&&\times e^{-m^{\prime 2}/M_{1}^{2}}e^{-m_{\Xi _{b}}^{2}/M_{2}^{2}}\left\{ %
\slashed q\slashed p\gamma _{5}-m_{\Xi _{b}}\slashed q\gamma _{5}\right.
\notag \\
&&\left. +\left( m^{\prime }-m_{\Xi _{b}}\right) \slashed p\gamma _{5}+\left[
m_{K}^{2}-m^{\prime }(m^{\prime }-m_{\Xi _{b}})\right] \gamma _{5}\right\} ,
\notag \\
&&  \label{eq:CFunc1/2}
\end{eqnarray}%
where $M_{1}^{2}$ and $M_{2}^{2}$ are the Borel parameters.

The QCD representation of the correlation function $\Pi ^{\mathrm{OPE}}(p,q)$
can be obtained by contracting the $s$ and $b$-quark fields, and inserting
relevant propagators into the obtained formulas. The explicit expressions of
the light-cone propagators of \ quarks are well known, and can be found, for
example, in Appendix of Ref.\ \cite{Agaev:2017lip}. After these operations
one gets formulas with matrix elements of non-local operators sandwiched
between the $K$-meson and vacuum states. The non-local quark operators emerge
and take their standard form after expansion of $\overline{s}_{\alpha
}^{a}u_{\beta }^{b}$ over full set of Dirac matrices $\Gamma ^{i}$
\begin{equation*}
\overline{s}_{\alpha }^{a}u_{\beta }^{b}=\frac{1}{4}\Gamma _{\beta \alpha
}^{i}(\overline{s}^{a}\Gamma ^{i}u^{b}),
\end{equation*}%
where $\Gamma ^{i}=1,\ \gamma _{5},\ \gamma _{\mu },\ i\gamma _{5}\gamma
_{\mu },\ \sigma _{\mu \nu }/\sqrt{2}$. The non-local quark-gluon operators
appear due to insertion of the gluon field strength tensor $G_{\lambda \rho
}(uv)$ from quark propagators into $\overline{s}_{\alpha }^{a}u_{\beta }^{b}$%
. These non-local quark and quark-gluon operators taken between the $K$
meson and vacuum generate $K$-meson's distribution amplitudes (DAs) of
various quark-gluon contents and twists.

\begin{figure}[h]
\includegraphics[width=10cm]{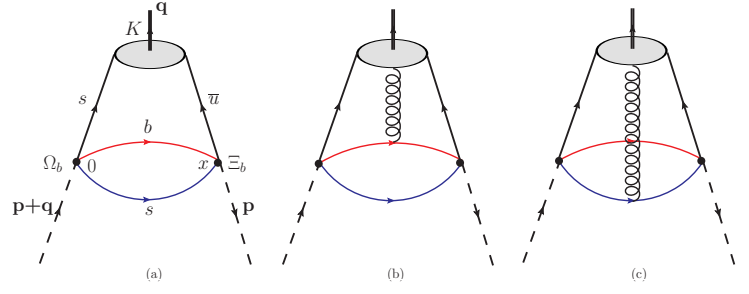}
\caption{Contributions to $\Pi ^{\mathrm{OPE}}(p,q)$ determined by
two-particle (a), and three-particle distribution amplitudes of $K$ meson
(b) and (c).}
\label{fig:Fig1}
\end{figure}

\begin{figure}[h]
\includegraphics[width=10cm]{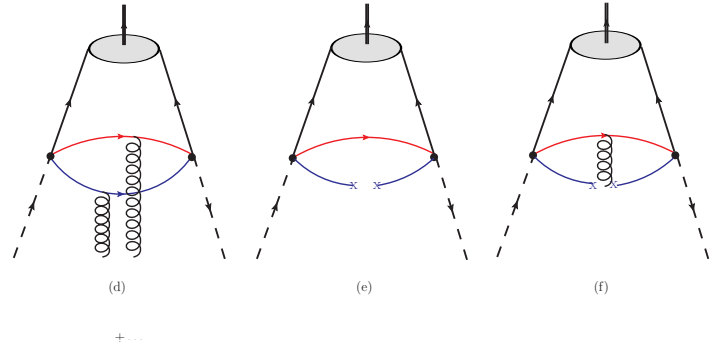}
\caption{Diagrams with gluon (a), quark (b), and mixed (c) vacuum
condensates.}
\label{fig:Fig2}
\end{figure}

Obtained contributions can be graphically represented by Feynman diagrams
some of which are plotted in Figs. \ref{fig:Fig1} and \ref{fig:Fig2}. The
leading order contribution is due to the diagram depicted in Fig.\ \ref%
{fig:Fig1} (a), which describes the perturbative term, where all of the
propagators are replaced by their perturbative components. Contribution of
this diagram can be found using the $K$-meson two particle distribution
amplitudes of two and higher twists. Components $\sim G_{\lambda \rho }$ in
one of the propagators lead to diagrams drawn in Figs.\ \ref{fig:Fig1} (b)
and (c). They are expressible in terms of three-particle DAs of $K$ meson.
There are also contributions to $\Pi ^{\mathrm{OPE}}(p,q)$ due to gluon,
quark and mixed vacuum condensates:  we demonstrate some of them in Figs.\ %
\ref{fig:Fig2} (a), (b) and (c), respectively.

The sum rules for the strong couplings can be derived after continuum
subtraction. There are two known approaches to perform this procedure. Thus,
in the context of the first method one calculates a double spectral density $%
\rho ^{\mathrm{OPE}}(s_{1},\,s_{2})$ as an imaginary part of the correlation
function, and using ideas of the quark-hadron duality carries out
subtraction. In the second approach it is necessary to get spectral density $%
\rho (s_{1},\,s_{2})$ directly from Borel transformation of the correlation
function in accordance with prescriptions developed in Refs.\ \cite%
{Ball:1994,Belyaev:1994zk,Aliev:2010yx,Aliev:2011ufa}. In this approach for $%
M_{1}^{2}=M_{2}^{2}=2M^{2}$ and $u_{0}=1/2$ (see, text below) the continuum
subtraction can be done using simple operations. For example, in the Borel
transformation of the correlation function terms
\begin{equation}
\left( M^{2}\right) ^{N}e^{-m^{2}/M^{2}}
\end{equation}%
preserve their original form if $N\leq 0$, and should be replaced by
\begin{equation}
\left( M^{2}\right) ^{N}e^{-m^{2}/M^{2}}\rightarrow \frac{1}{\Gamma (N)}%
\int_{m^{2}}^{s_{0}}dse^{-s/M^{2}}\left( s-m^{2}\right) ^{N-1},
\end{equation}%
if $N>0$. The subtracted version of other expressions, which emerge in
calculations are collected in the Appendix \ref{sec:App}. In the present work to
perform the continuum subtraction we follow these procedures.

To derive the sum rules for the strong couplings it is possible to use
different Lorentz structures in Eq.\ (\ref{eq:CFunc1/2}). We have found that
structures $\sim \slashed q\slashed p\gamma _{5}$ and $\sim \slashed p\gamma
_{5}$ are convenient for our purposes. Isolating the corresponding terms in the
Borel transformed form of the correlation function $\Pi ^{\mathrm{OPE}}(p,\,q)$
we obtain:
\begin{equation}
g_{\widetilde{\Omega }_{b}\Xi _{b}K}=\frac{e^{\widetilde{m}%
^{2}/M_{1}^{2}}e^{m_{\Xi _{b}}^{2}/M_{2}^{2}}}{\lambda _{\Xi _{b}}\widetilde{%
\lambda }(m^{\prime }+\widetilde{m})}\left[ (m^{\prime }-m_{\Xi _{b}})%
\mathcal{B}\Pi _{1}^{\mathrm{OPE}}-\mathcal{B}\Pi _{2}^{\mathrm{OPE}}\right]
,
\end{equation}%
and
\begin{equation}
g_{\Omega _{b}^{\prime }\Xi _{b}K}=\frac{e^{m^{\prime 2}/M_{1}^{2}}e^{m_{\Xi
_{b}}^{2}/M_{2}^{2}}}{\lambda _{\Xi _{b}}\lambda ^{\prime }(m^{\prime }+%
\widetilde{m})}\left[ (\widetilde{m}+m_{\Xi _{b}})\mathcal{B}\Pi _{1}^{%
\mathrm{OPE}}+\mathcal{B}\Pi _{2}^{\mathrm{OPE}}\right] ,
\end{equation}%
where $\Pi _{1}^{\mathrm{OPE}}(p^{2},\,p^{\prime 2})$ and $\Pi _{2}^{\mathrm{%
OPE}}(p^{2},\,p^{\prime 2})$ are the invariant amplitudes corresponding to
the structures $\slashed q\slashed p\gamma _{5}$ and $\slashed p\gamma _{5}$,
respectively.

Because the masses of the initial $\Omega _{b}$ and final $\Xi _{b}^{0}$
baryons are close to each other we choose $M_{1}^{2}=M_{2}^{2}$, and
introduce the Borel parameter $M^{2}$ through the equality%
\begin{equation}
\frac{1}{M^{2}}=\frac{1}{M_{1}^{2}}+\frac{1}{M_{2}^{2}},
\end{equation}%
which simplifies considerably the obtained expressions. In the Appendix we write
down the full expression for $\mathcal{B}\Pi _{1}^{\mathrm{OPE}}=$ $\Pi
_{1}(M^{2})$ in terms of $K$-meson's DAs. Some of $K$ meson DAs and values
of corresponding parameters are also collected there.

Using the couplings $g_{\widetilde{\Omega }_{b}\Xi _{b}K}$ and $g_{\Omega
_{b}^{\prime }\Xi _{b}K}$ it is not difficult to calculate the width of $%
\widetilde{\Omega }_{b}^{-}\rightarrow \Xi _{b}^{0}K^{-}$ and $\Omega
_{b}^{\prime -}\rightarrow \Xi _{b}^{0}K^{-}$ decays. The required
expressions are presented below:
\begin{eqnarray}
\Gamma \left( \widetilde{\Omega }_{b}\rightarrow \Xi _{b}^{0}K^{-}\right) &=&%
\frac{g_{\widetilde{\Omega }_{b}\Xi _{b}K}^{2}}{8\pi \widetilde{m}^{2}}\left[
(\widetilde{m}+m_{\Xi _{b}})^{2}-m_{K}^{2}\right]  \notag \\
&&\times f(\widetilde{m},m_{\Xi _{b}},m_{K}).
\end{eqnarray}%
and%
\begin{eqnarray}
\Gamma \left( \Omega _{b}^{\prime }\rightarrow \Xi _{b}^{0}K^{-}\right) &=&%
\frac{g_{\Omega _{b}^{\prime }\Xi _{b}K}^{2}}{8\pi m^{\prime 2}}\left[
(m^{\prime }-m_{\Xi _{b}})^{2}-m_{K}^{2}\right]  \notag \\
&&\times f(m^{\prime },m_{\Xi _{b}},m_{K}),
\end{eqnarray}%
In expressions above the function $f(x,y,z)$ is given as:
\begin{equation*}
f(x,y,z)=\frac{1}{2x}\sqrt{%
x^{4}+y^{4}+z^{4}-2x^{2}y^{2}-2x^{2}z^{2}-2y^{2}z^{2}}.
\end{equation*}


\subsection{Decays $\widetilde{\Omega}_b^{\star} \to \Xi_{b}^{0}K^{-}$ and $%
\Omega_b^{\star \prime} \to \Xi_{b}^{0}K^{-}$}

The decays of the spin-$3/2$ baryons $\widetilde{\Omega }_{b}^{\star }$ and $%
\Omega _{b}^{\star \prime }$ to $\Xi _{b}^{0}\,K^{-}$ can be analyzed as it
has been done in previous subsection for the spin-$1/2$ baryons. To this
end, we consider the correlation function
\begin{equation}
\Pi _{\mu }(p,q)=i\int d^{4}xe^{ipx}\langle K(q)|\mathcal{T}\{\eta _{\Xi
_{b}}(x)\overline{\eta }_{\mu }(0)\}|0\rangle ,
\end{equation}%
where the interpolating current $\eta _{\mu }(x)$ is given in the form
\begin{equation}
\eta _{\mu }=\frac{1}{\sqrt{3}}\epsilon ^{abc}\left[ \left( s^{aT}C\gamma
_{\mu }s^{b}\right) b^{c}+2\left( s^{aT}C\gamma _{\mu }b^{b}\right) s^{c}%
\right] .  \label{eq:BayC3/2}
\end{equation}

In order to express the function $\Pi _{\mu }(p,q)$ in terms of the physical
parameters of the involved particles we follow the same manipulations as in the
case of the spin-1/2 baryons, the difference being only in definitions of the
relevant matrix elements. Thus, we employ the following matrix elements for
the spin-3/2 baryons
\begin{eqnarray}
\langle 0|\eta _{\mu }|\widetilde{\Omega }_{b}^{\star }(p,s)\rangle &=&%
\widetilde{\lambda }^{\star }\gamma _{5}\widetilde{u}_{\mu }(p,s),  \notag \\
\langle 0|\eta _{\mu }|\Omega _{b}^{\star \prime }(p,s)\rangle &=&\lambda
^{\star \prime }u_{\mu }^{\prime }(p,s),  \label{eq:MElem2}
\end{eqnarray}%
where $u_{\mu }(p,s)$ are Rarita-Schwinger spinors, and $\widetilde{\lambda }%
^{\star }$ and $\lambda ^{\star \prime }$ are residues of the  $\widetilde{\Omega
}_{b}^{\star }$ and $\Omega _{b}^{\star \prime }$ baryons, respectively.

We introduce also the strong couplings $g_{\widetilde{\Omega }_{b}^{\star
}\Xi _{b}K}$ and $g_{\Omega _{b}^{\star \prime }\Xi _{b}K}$ by means of the
formulas
\begin{eqnarray}
&&\langle K(q)\Xi _{b}^{0}(p,s)|\widetilde{\Omega }_{b}^{\star }(p^{\prime
},s^{\prime })\rangle =g_{\widetilde{\Omega }_{b}^{\star }\Xi _{b}K}%
\overline{u}(p,s)\gamma _{5}u_{\alpha }(p^{\prime },s^{\prime })q^{\alpha },
\notag \\
&&\langle K(q)\Xi _{b}^{0}(p,s)|\Omega _{b}^{\star \prime }(p^{\prime
},s^{\prime })\rangle =g_{\Omega _{b}^{\star \prime }\Xi _{b}K}\overline{u}%
(p,s)u_{\alpha }(p^{\prime },s^{\prime })q^{\alpha }.  \notag \\
&&{}  \label{eq:MElem3}
\end{eqnarray}

Substituting the matrix elements given by Eqs.\ (\ref{eq:MElem2}) and (\ref%
{eq:MElem3}) into $\Pi _{\mu }^{\mathrm{Phys}}(p,q)$ and performing the
summation over the spins  in accordance with the expression
\begin{eqnarray}
&&\sum\limits_{s}u_{\mu }(p,s)\overline{u}_{\nu }(p,s)=-(\slashed p+m)\left[
g_{\mu \nu }-\frac{1}{3}\gamma _{\mu }\gamma _{\nu }\right.  \notag \\
&&\left. -\frac{2}{3m^{2}}p_{\mu }p_{\nu }+\frac{1}{3m}(p_{\mu }\gamma _{\nu
}-p_{\nu }\gamma _{\mu })\right] ,
\end{eqnarray}%
we get
\begin{eqnarray}
&&\Pi _{\mu }^{\mathrm{Phys}}(p,q)=\frac{g_{\widetilde{\Omega }_{b}^{\star
}\Xi _{b}K}\lambda _{\Xi _{b}}\widetilde{\lambda }^{\star }}{(p^{2}-m_{\Xi
_{b}}^{2})(p^{\prime }{}^{2}-\widetilde{m}^{\star 2})}q^{\alpha }(\slashed %
p+m_{\Xi _{b}})\gamma _{5}  \notag \\
&&\times \left( \slashed p+\slashed q+\widetilde{m}^{\star }\right)
F_{\alpha \mu }(\widetilde{m}^{\star })\gamma _{5}  \notag \\
&&-\frac{g_{\Omega _{b}^{\star \prime }\Xi _{b}K}\lambda _{\Xi _{b}}\lambda
^{\ast \prime }}{(p^{2}-m_{\Xi _{c}}^{2})(p^{\prime 2}-m^{\prime 2})}%
q^{\alpha }(\slashed p+m_{\Xi _{b}})  \notag \\
&&\times \left( \slashed p+\slashed q+m^{\ast \prime }\right) F_{\alpha \mu
}(m^{\ast \prime })+\ldots.  \label{eq:CFspin3/2}
\end{eqnarray}%
In Eq.\ (\ref{eq:CFspin3/2}) we have used the notation
\begin{eqnarray}
&&F_{\alpha \mu }(m)=g_{\alpha \mu }-\frac{1}{3}\gamma _{\alpha }\gamma
_{\mu }-\frac{2}{3m^{2}}(p_{\alpha }+q_{\alpha })(p_{\mu }+q_{\mu })  \notag
\\
&&+\frac{1}{3m}\left[ (p_{\alpha }+q_{\alpha })\gamma _{\mu }-(p_{\mu
}+q_{\mu })\gamma _{\alpha }\right] .
\end{eqnarray}%
For the Borel transformation of $\Pi _{\mu }^{\mathrm{Phys}}(p,q)$ we obtain%
\begin{eqnarray}
&&\mathcal{B}\Pi _{\mu }^{\mathrm{Phys}}(p,q)=g_{\widetilde{\Omega }%
_{b}^{\star }\Xi _{b}K}\lambda _{\Xi _{b}}\widetilde{\lambda }e^{-\widetilde{%
m}^{2}/M_{1}^{2}}e^{-m_{\Xi _{b}}^{2}/M_{2}^{2}}q^{\alpha }  \notag \\
&&\times (\slashed p+m_{\Xi _{b}})\gamma _{5}\left( \slashed p+\slashed q+%
\widetilde{m}^{\star }\right) F_{\alpha \mu }(\widetilde{m}^{\star })\gamma
_{5}  \notag \\
&&-g_{\Omega _{b}^{\star \prime }\Xi _{b}K}\lambda _{\Xi _{b}}\lambda
^{\star \prime }e^{-m^{\prime 2}/M_{1}^{2}}e^{-m_{\Xi
_{b}}^{2}/M_{2}^{2}}q^{\alpha }(\slashed p+m_{\Xi _{b}})  \notag \\
&&\times \left( \slashed p+\slashed q+m^{\star \prime }\right) F_{\alpha \mu
}(m^{\star \prime }).
\end{eqnarray}%
The required sum rules can be obtained by using invariant amplitudes
corresponding to the structures $\slashed q\slashed p\gamma _{\mu }$ and $%
\slashed qq_{\mu }$.

The correlation function $\Pi _{\mu }^{\mathrm{OPE}}(p,q)$ is determined in
terms of numerous distribution amplitudes of the $K$ meson. In Appendix we
also provide the explicit expression for double Borel transformed form  of the
invariant amplitude corresponding to the structure $\slashed{q}\slashed{p}%
\gamma _{\mu }$ $.$ By fixing the same structures in both $\mathcal{B}\Pi
_{\mu }^{\mathrm{Phys}}(p,q)$ and $\mathcal{B}\Pi _{\mu }^{\mathrm{OPE}%
}(p,q) $ and equating Borel transformed form of the relevant invariant
amplitudes, it is possible to get and solve two equations for the strong
couplings $g_{\widetilde{\Omega }_{b}^{\star }\Xi _{b}K}$ and $g_{\Omega
_{b}^{\star \prime }\Xi _{b}K}$.

Then the width of the $\widetilde{\Omega }_{b}^{\star }\rightarrow \Xi
_{b}^{0}K^{-}$ decay can be obtained as%
\begin{eqnarray}
\Gamma (\widetilde{\Omega }_{b}^{\star } &\rightarrow &\Xi _{b}^{0}K^{-})=%
\frac{g_{\widetilde{\Omega }_{b}^{\star }\Xi _{b}K}^{2}}{24\pi \widetilde{m}%
^{\star 2}}\left[ (\widetilde{m}^{\star }-m_{\Xi _{b}})^{2}-m_{K}^{2}\right]
\notag \\
&&\times f^{3}(\widetilde{m}^{\star },m_{\Xi _{b}},m_{K}),
\end{eqnarray}%
whereas for $\Gamma (\Omega _{c}^{\star \prime }\rightarrow \Xi
_{b}^{0}K^{-})$ we find
\begin{eqnarray}
\Gamma (\Omega _{b}^{\star \prime } &\rightarrow &\Xi _{b}^{0}K^{-})=\frac{%
g_{\Omega _{b}^{\star \prime }\Xi _{b}K}^{2}}{24\pi m^{\star \prime 2}}\left[
(m^{\star \prime }+m_{\Xi _{b}})^{2}-m_{K}^{2}\right]  \notag \\
&&\times f^{3}(m^{\star \prime },m_{\Xi _{b}},m_{K}).
\end{eqnarray}%
These expressions will be used in  numerical
calculations.


\section{Numerical computations}

\label{sec:NumRest} 
The obtained sum rules for the strong couplings depend on numerous parameters.
First of all, the light-cone propagator of $s-$quark contains the quark and
mixed vacuum condensates numerical value of which $\langle \overline{s}%
s\rangle =-0.8\times (0.24\pm 0.01)^{3}~\mathrm{GeV}^{3}$, $\langle
\overline{s}g_{s}\sigma Gs\rangle =m_{0}^{2}\langle \overline{s}s\rangle $,
where $m_{0}^{2}=(0.8\pm 0.1)~\mathrm{GeV}^{2}$ are well known. For the
gluon condensate we utilize $\langle \alpha _{s}G^{2}/\pi \rangle =(0.012\pm
0.004)~\mathrm{GeV}^{4}$. The masses of the $b-$ and $s$-quarks are
presented in PDG \cite{Olive:2016xmw}: $m_{b}=4.18_{-0.03}^{+0.04}~\mathrm{%
GeV}$ and $m_{s}=96_{-4}^{+8}~\mathrm{MeV}$. The residue $\lambda _{\Xi
_{b}}=0.054\pm 0.012$ $\mathrm{GeV}^{3}$ of $\Xi _{b}^{0}$ baryon is
borrowed from Ref.\ \cite{Azizi:2016dmr}.

Calculations within the sum rule method imply fixing of the working windows
for the Borel parameter $M^{2}$ and continuum threshold $s_{0}$, which are
two auxiliary parameters of computations. In addition, formulas for the spin-%
$1/2$ baryons depend on $\beta $ arising from the expressions of
the interpolating currents $\eta (x)$ and $\eta_{\Xi_b}(x)$.  The
mass and pole residue of the excited bottom baryons also appear in
the sum rules for the strong couplings as input parameters. In our
previous work \cite{Agaev:2017jyt} we evaluated the spectroscopic parameters of the $%
\widetilde{\Omega }_{b}$, $\Omega _{b}^{\prime }$ and $\widetilde{\Omega }%
_{b}^{\star }$, $\Omega _{b}^{\star \prime }$ baryons. Predictions
obtained there for the mass and pole residue of $1P$ and $2S$
bottom baryons with $J=1/2$ and $J=3/2$, as well as the working
ranges of the parameters $M^{2}$ and $s_{0}$ are collected in
Table \ref{tab:MassRes}. Results for the spin-$1/2$ baryons were
extracted by varying the parameter $\beta =\tan \theta $ in Eq.\
(\ref{eq:BayC1/2}) within the limits
\begin{equation}
-0.75\leq \cos \theta \leq -0.45,\ 0.45\leq \cos \theta \leq 0.75,
\label{eq:Beta}
\end{equation}
which led to stable predictions for their masses and residues.

The choice of $M^{2}$, $s_{0}$  and $\beta $ is not arbitrary, but
has to satisfy restrictions of sum rule calculations. Thus, the
upper bound of the working region for $M^2$ is obtained from the
constraint imposed on the pole contribution
\begin{equation}
\frac{\Pi^{\mathrm{OPE}}(M^2,\ s_0, \
\beta)}{\Pi^{\mathrm{OPE}}(M^2,\ \infty, \ \beta)} > \frac{1}{2},
\label{eq:Rest1}
\end{equation}
where  $\Pi^{\mathrm{OPE}}(M^2,\ s_0,\ \beta)$ is the Borel
transformation of the relevant correlation function after
continuum subtraction.

The lower limit of the Borel parameter $M^2$ is determined from
exceeding of the perturbative contribution over the
nonperturbative one as well as convergence of the operator product
expansion. In the present work we apply the following criteria: at
the lower bound of the Borel window the perturbative contribution
has to constitute $\geq 80 \% $ part of the corresponding sum
rule, and contribution of the highest dimensional term (i.e., in
our case $\mathrm{Dim}9$ term ) should not exceed $1\% $ of the
whole result.

The limits within of which the parameter $s_0$ can be varied are
determined from the pole to total contribution ratio to achieve
its greatest possible value. Quantities extracted from sum rules
have also to demonstrate minimal dependence on $M^2$ while varying
$s_0$ in the allowed domain.
\begin{widetext}

\begin{table}[tbp]
\begin{tabular}{|c|c|c|c|c|}
\hline
$(n,J^{P})$ & $(1P,\frac{1}{2}^{-})$ & $(2S,\frac{1}{2}^{+})$ & $(1P,\frac{3%
}{2}^{-})$ & $(2S,\frac{3}{2}^{+})$   \\ \hline $M^2
~(\mathrm{GeV}^2$) & $6.5-9.5$ & $6.5-9.5$ & $6.5-9.5$ & $6.5-9.5$
\\ \hline
$s_0 ~(\mathrm{GeV}^2$) & $6.6^2-6.8^2$ & $6.8^2-7.0^2$ & $6.7^2-6.9^2$ & $%
6.9^2-7.1^2$   \\ \hline $m_{\Omega_b} ~(\mathrm{MeV})$ & $6336
\pm 183$ & $6487 \pm 187$ & $6301 \pm 193$ & $6422 \pm 198$
\\ \hline
$\lambda_{\Omega_b}\cdot 10^{2} ~(\mathrm{GeV}^3)$ & $17.5 \pm 2.9$ & $%
19.8\pm 4.1$ & $19.2 \pm 3.1$ & $29.1 \pm 5.3$   \\ \hline
\end{tabular}%
\caption{The $m_{\Omega _{b}}$ and $\protect\lambda _{\Omega
_{b}}$ of the excited bottom baryons with $J=1/2$ and $J=3/2$.}
\label{tab:MassRes}
\end{table}

\end{widetext}
Finally, we  determine a working range for $\beta $ by demanding a
weak dependence of our results on its choice, which quantitatively
reads
\begin{equation}
\frac{|\Pi^{\mathrm{OPE}}(M^2,\ s_0, \beta_0
)-\Pi^{\mathrm{OPE}}(M^2,\ s_0, \beta_0 \pm \Delta \beta
)|}{\Pi^{\mathrm{OPE}}(M^2,\  s_0, \ \beta_0)} \leq 0.1,
\label{eq:Rest2}
\end{equation}
where $\beta_0 \pm \Delta \beta \in [\beta_{\mathrm{min}}, \
\beta_{\mathrm{max}}]$.

In the choice of the regions for  $M^{2}$, $s_{0}$, and $\beta $
we keep in mind that sum rules for masses and pole residues  of
the excited $\Omega_b$ baryons also depend on these parameters.
Because they enter as input quantities to sum rules for the strong
couplings a deviation from regions found in Ref.\
\cite{Agaev:2017jyt} may generate additional uncertainties.

Analysis carried out in accordance with these requirements enables
us to fix the parameters $M^{2}$, $s_{0}$ and $\beta$. Thus, for
both the spin-1/2 and spin-3/2 bottom baryons the working region
for the
Borel parameter is%
\begin{equation*}
M^{2}\in \lbrack 6.5-9.5]\ \mathrm{GeV}^{2}.
\end{equation*}%
The regions for the continuum threshold $s_{0}$ depend on type of the $%
\Omega _{b}$ baryon under consideration. For calculation of the strong
coupling of $1P$ and $2S$ excitations of the spin-1/2 baryon we use
\begin{eqnarray}
s_{0} &\in &[6.6^{2}-6.8^{2}]\ \mathrm{GeV}^{2},  \notag \\
s_{0} &\in &[6.8^{2}-7.0^{2}]\ \mathrm{GeV}^{2},  \label{eq:WR1}
\end{eqnarray}%
respectively. For the same excited states of the spin-3/2 baryon we get
\begin{eqnarray}
s_{0} &\in &[6.7^{2}-6.9^{2}]\ \mathrm{GeV}^{2},  \notag \\
s_{0} &\in &[6.9^{2}-7.1^{2}]\ \mathrm{GeV}^{2}.  \label{eq:WR2}
\end{eqnarray}%
 For spin-1/2 particles the parameter $\beta $ is fixed as in Eq.\
(\ref{eq:Beta}).

In regions chosen for $M^{2}$,  $s_{0}$ and $\beta$ the sum rules
comply aforementioned constraints. Thus,
in Fig.\ \ref{fig:PC} we plot the pole contribution to the sum rule for $%
g_{\widetilde{\Omega }_{b}^{\star }\Xi _{b}K}$, which  at
$M^2=9.5\ \mathrm{GeV}^2$ equals to $64\%$ of the whole
contribution, and reaches $75\%$ of its value in the case of
$g_{\Omega _{b}^{\star \prime }\Xi _{b}K}$.

In Fig.\ \ref{fig:PnonP} we compare  the perturbative and
nonperturbative contributions to the strong coupling
$g_{\widetilde \Omega_b^{\star }\Xi_{b}K }$ as functions of $M^2$
and $s_0$ at central values of $s_0$ and $M^2$, respectively. It
is seen, that the perturbative contribution amounts to more than
$0.8$ part of the result. Convergence of OPE becomes evident from
analysis of Fig.\ \ref{fig:Dim}, where by the curve labelled $\geq
\mathrm{Dim}6$ we depict the sum of nonperturbative terms from
sixth till ninth dimensions. They already satisfy the imposed
constraint on nonperturbative terms to guaranty convergence of the
expansion.

Dependence on $\beta $ is mild: at the
central values of $M^{2}=8$ $\mathrm{GeV}^{2}$ and $s_{0}=6.7^{2}\ \mathrm{%
GeV}^{2}$ variation of $\beta $ within limits determined by Eq.\
(\ref{eq:Beta}) leads only to $\sim 7\%$ \ changes in
$g_{\widetilde{\Omega }_{b}\Xi _{b}K}$, whereas at $M^{2}=8$
$\mathrm{GeV}^{2}$ and $s_{0}=6.9^{2}\ \mathrm{GeV}^{2}$ they
amount approximately to $8\%$ of $g_{\Omega _{b}^{\prime }\Xi
_{b}K}$. In the whole region of $M^{2}$ and $s_{0}$ they do not
overshoot $10\%$ of the results, and are in agreement with Eq.\
(\ref{eq:Rest2}).
\begin{widetext}

\begin{figure}[h!]
\begin{center}
\includegraphics[totalheight=6cm,width=8cm]{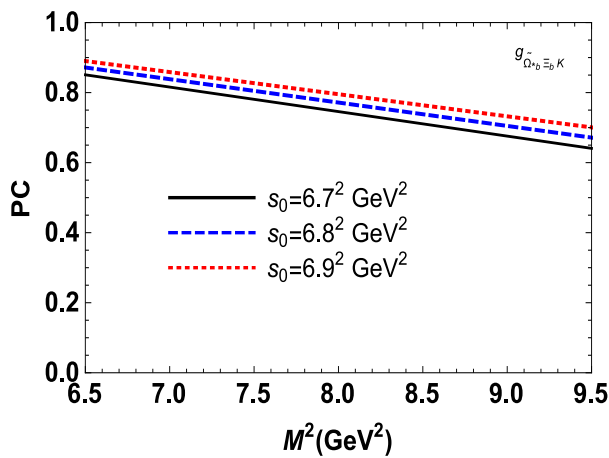}
\includegraphics[totalheight=6cm,width=8cm]{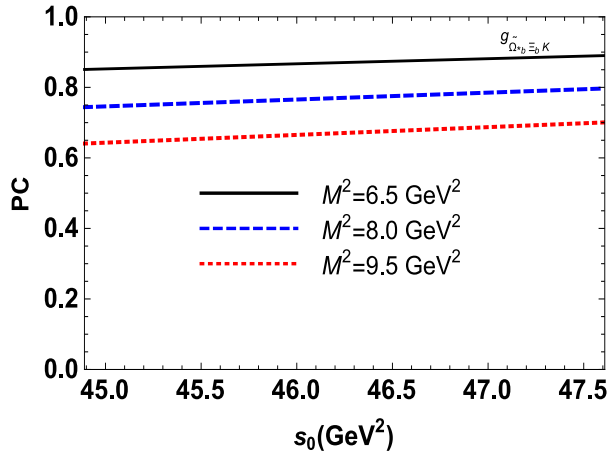}
\end{center}
\caption{ The dependence of the pole contribution to
$g_{\widetilde \Omega_b^{\star }\Xi_{b}K }$ on the Borel parameter
$M^2$ (left panel), and on the continuum threshold $s_0$ (right
panel).} \label{fig:PC}
\end{figure}
\begin{figure}[h!]
\begin{center}
\includegraphics[totalheight=6cm,width=8cm]{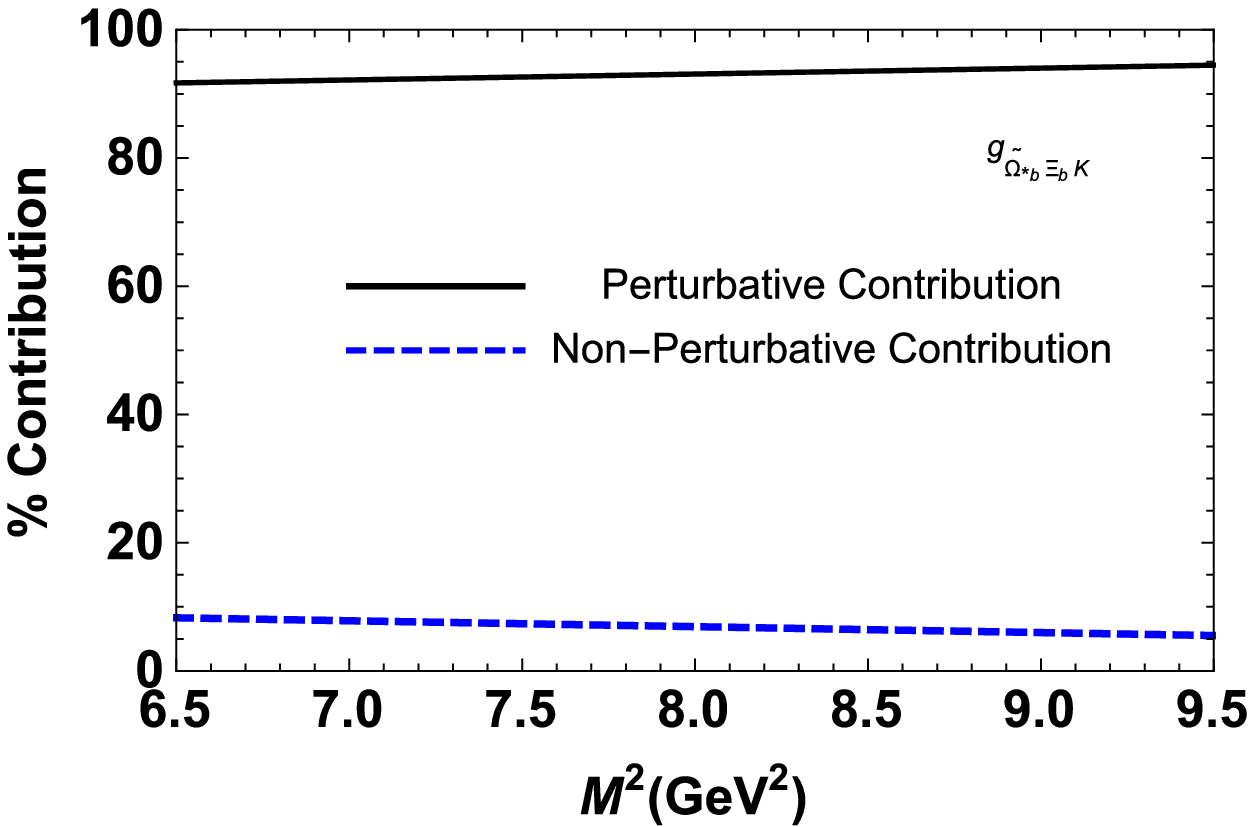}
\includegraphics[totalheight=6cm,width=8cm]{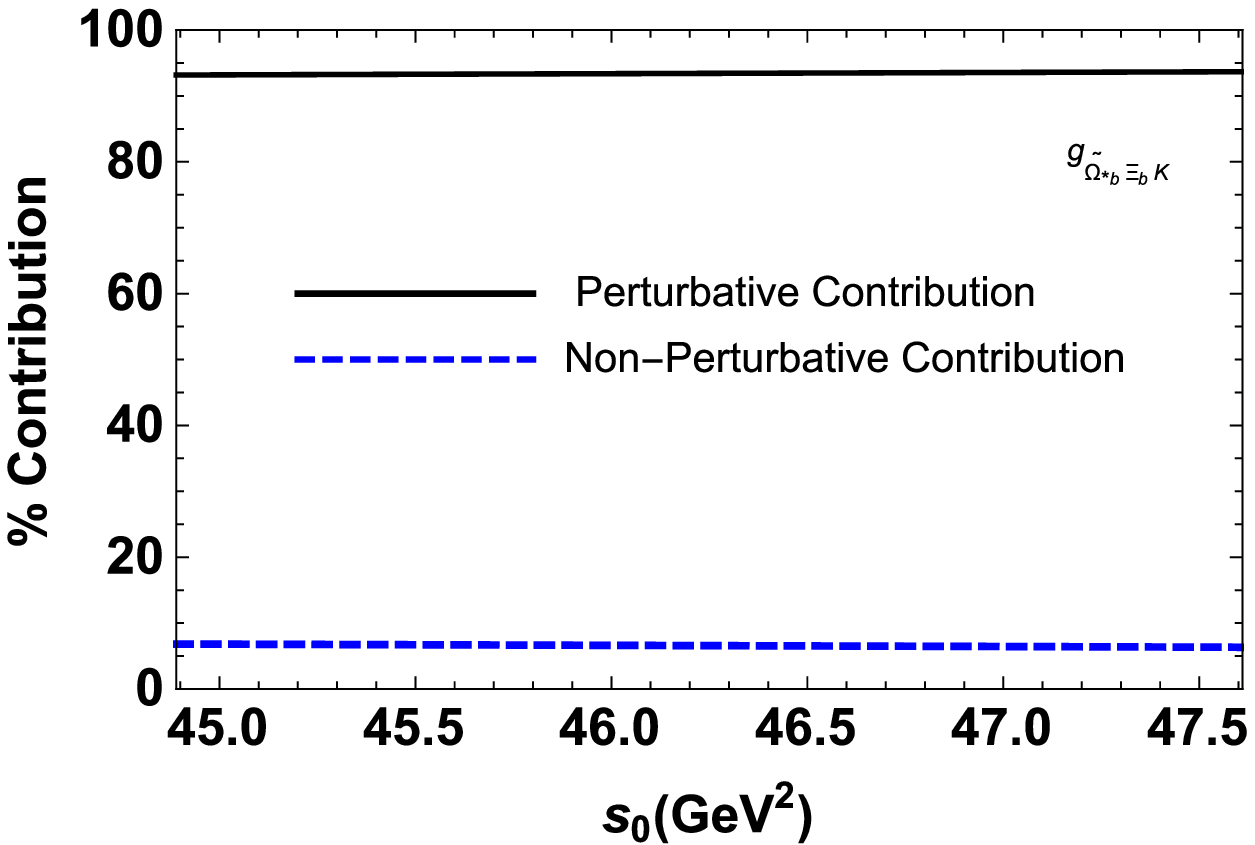}
\end{center}
\caption{ The perturbative and nonperturbative contributions to
the coupling $g_{\widetilde \Omega_b^{\star }\Xi_{b}K }$  as
functions of the Borel parameter $M^2$ (left panel), and of the
continuum threshold $s_0$ (right panel).} \label{fig:PnonP}
\end{figure}
\begin{figure}[h!]
\begin{center}
\includegraphics[totalheight=6cm,width=8cm]{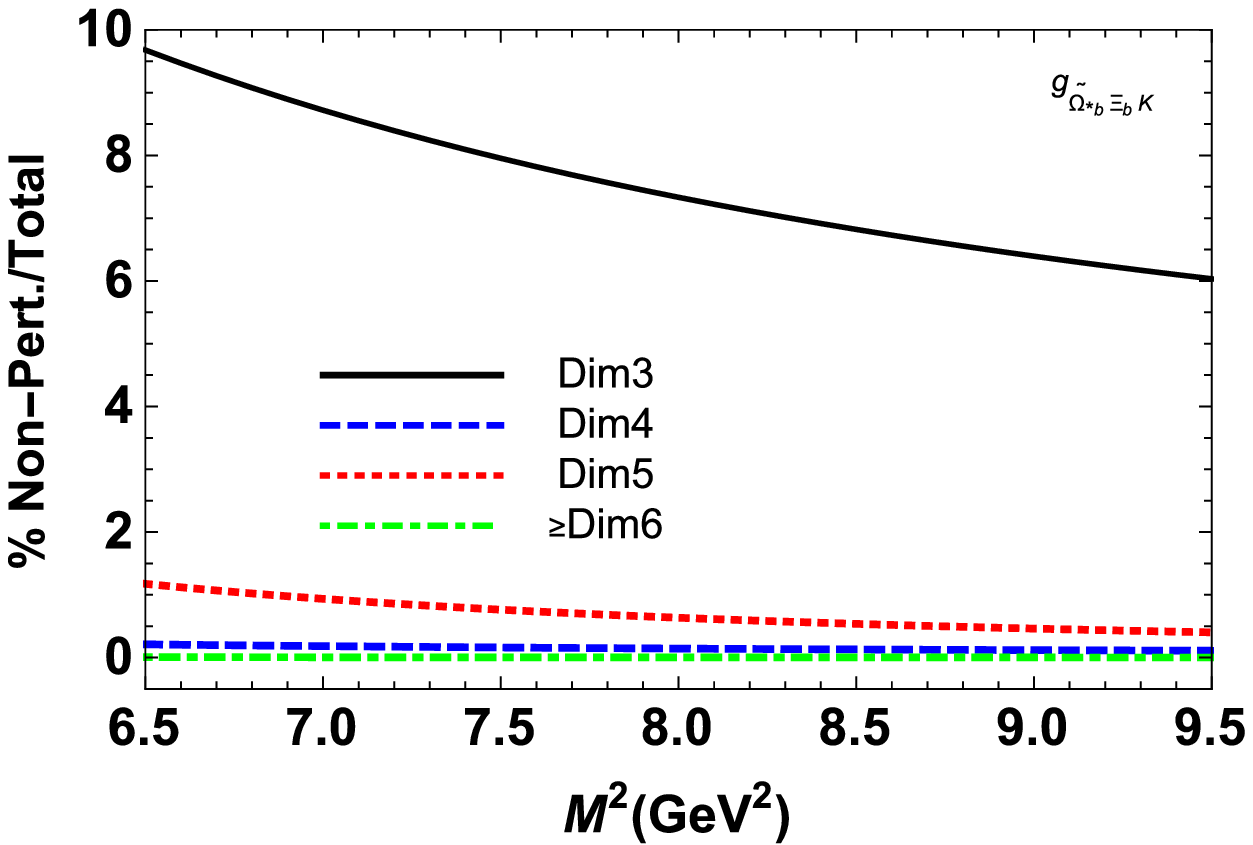}
\includegraphics[totalheight=6cm,width=8cm]{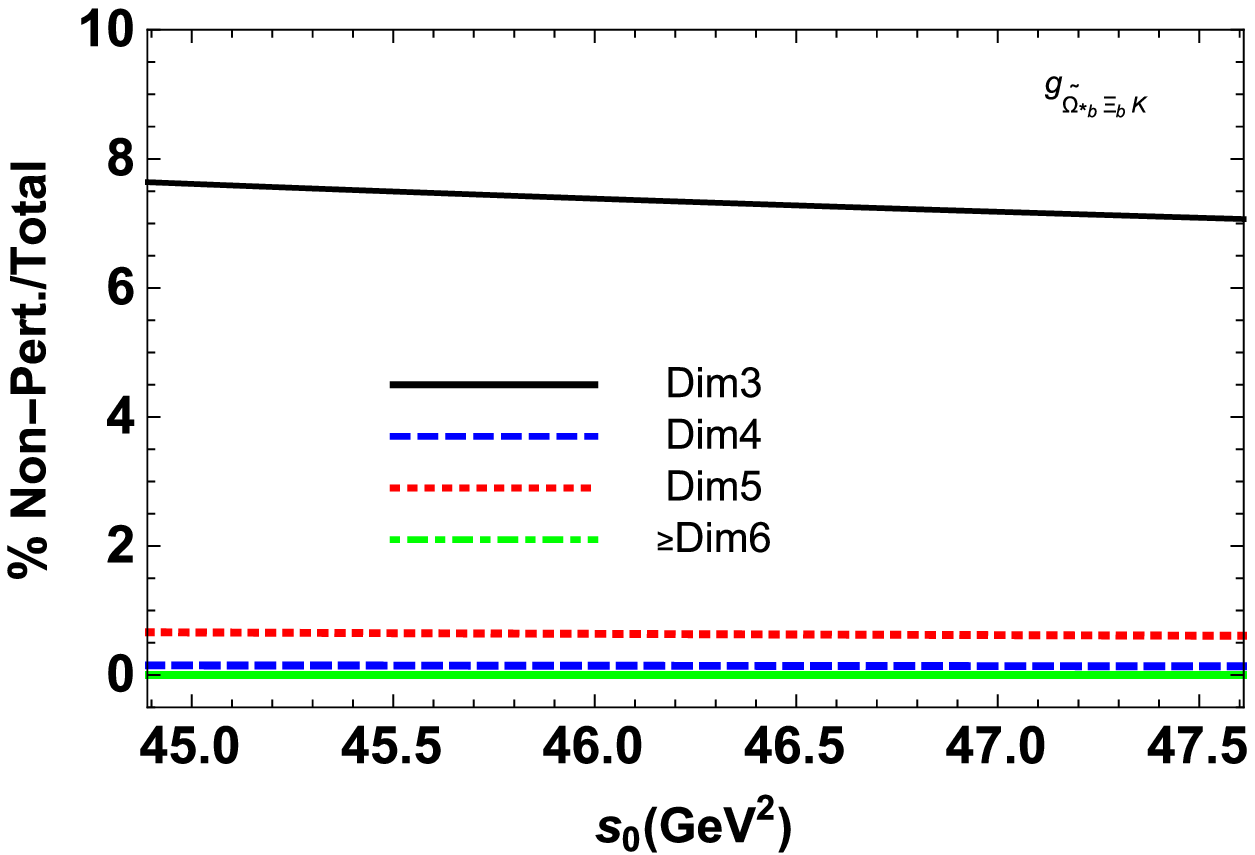}
\end{center}
\caption{ The nonperturbative contributions to the strong coupling
$g_{\widetilde \Omega_b^{\star }\Xi_{b}K }$ as functions of the
Borel parameter $M^2$ (at $s_0=46.25\  \mathrm{GeV}^2$, left
panel), and of the continuum threshold $s_0$ ($M^2=8\
\mathrm{GeV}^2$, right panel).} \label{fig:Dim}
\end{figure}

\end{widetext}
The regions for $M^{2}$ and $s_{0}$ in the light-cone sum rule
computations of the strong couplings $g_{\widetilde{\Omega
}_{b}\Xi
_{b}K}$, $\ g_{\Omega _{b}^{\prime }\Xi _{b}K}$ , $g_{\widetilde{\Omega }%
_{b}^{\star }\Xi _{b}K}$ and $g_{\Omega _{b}^{\star \prime }\Xi
_{b}K}$
coincide with ones used in calculations of the mass and residue of  $%
\widetilde{\Omega }_{b}$, $\Omega _{b}^{\prime }$, $\widetilde{\Omega }%
_{b}^{\star }$ and $\Omega _{b}^{\star \prime }$ baryons. By such
choice of working windows for $M^2$, $s_0$ and $\beta$ we also
evade appearance of additional theoretical uncertainties.

The strong couplings of the excited spin-1/2 $\Omega _{b}$ baryons
equal to:
\begin{equation}
g_{\widetilde{\Omega }_{b}\Xi _{b}K}=0.36\pm 0.07,\ \ g_{\Omega _{b}^{\prime
}\Xi _{b}K}=7.33\pm 1.61.\
\end{equation}%
For couplings of the $\Omega _{b}^{\star }$ baryons we get
\begin{equation}
g_{\widetilde{\Omega }_{b}^{\star }\Xi _{b}K}=82.29\pm 14.08,\ g_{\Omega
_{b}^{\star \prime }\Xi _{b}K}=1.04\pm 0.28.\
\end{equation}%
Here we provide also theoretical errors of our predictions
essential part of
which comes from uncertainties in the choice of the auxiliary parameters $%
M^{2}$ and $s_{0}$ (for spin-1/2 baryons also from $\beta $).
Theoretical errors vary from $\pm 15\%$ \ for
$g_{\widetilde{\Omega }_{b}^{\star }\Xi _{b}K}$ till $\pm 27\%$
for $g_{\Omega _{b}^{\star \prime }\Xi _{b}K}$ and do not exceed
$30\%$ of the central values, which is an accuracy accepted in QCD
sum rule calculations. To demonstrate a sensitivity of the
obtained results to choice of these parameters in Figs.\
\ref{fig:Coupl1}, \ref{fig:Coupl1A} and \ref{fig:Coupl2} we plot
$g_{\Omega _{b}^{\prime }\Xi _{b}K}$, $g_{\widetilde\Omega
_{b}^{\star }\Xi _{b}K}$ and $g_{\Omega _{b}^{\star \prime }\Xi
_{b}K}$ as functions of $M^{2}$ at fixed $s_{0}$, and functions of
$s_{0}$ for chosen $M^{2}$.

For width of the excited $1P$ and $2S$ bottom baryons' decays we
find: for $\Omega _{b}$
\begin{eqnarray}
\Gamma \left( \widetilde{\Omega }_{b}\rightarrow \Xi _{b}^{0}K^{-}\right)
&=&3.97\pm 0.91\ \mathrm{MeV},\   \notag \\
\Gamma \left( \Omega _{b}^{\prime }\rightarrow \Xi
_{b}^{0}K^{-}\right) &=&5.51\pm 1.42\ \mathrm{MeV},
\label{eq:FRes1}
\end{eqnarray}%
and for $\Omega _{b}^{\star}$
\begin{eqnarray}
\ \Gamma (\widetilde{\Omega }_{b}^{\star } &\rightarrow &\Xi
_{b}^{0}K^{-})=0.04\pm 0.01\ \mathrm{MeV},  \notag \\
\Gamma (\Omega _{b}^{\star \prime } &\rightarrow &\Xi _{b}^{0}K^{-})=2.57\pm
0.78\ \mathrm{MeV}.  \label{eq:FRes2}
\end{eqnarray}%
The predictions for width of the decay processes given by Eqs.\ (\ref%
{eq:FRes1}) and (\ref{eq:FRes2}) are our final results.
\begin{widetext}

\begin{figure}[h!]
\begin{center}
\includegraphics[totalheight=6cm,width=8cm]{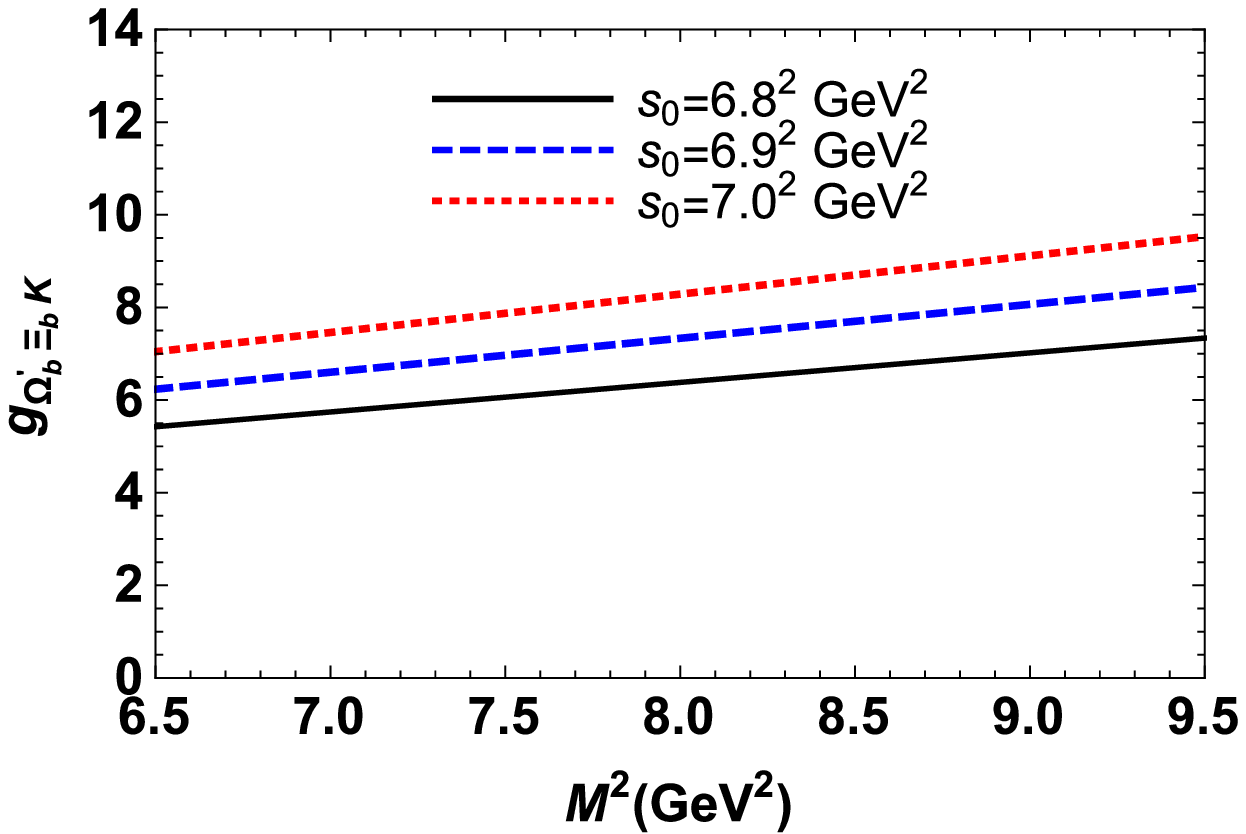}
\includegraphics[totalheight=6cm,width=8cm]{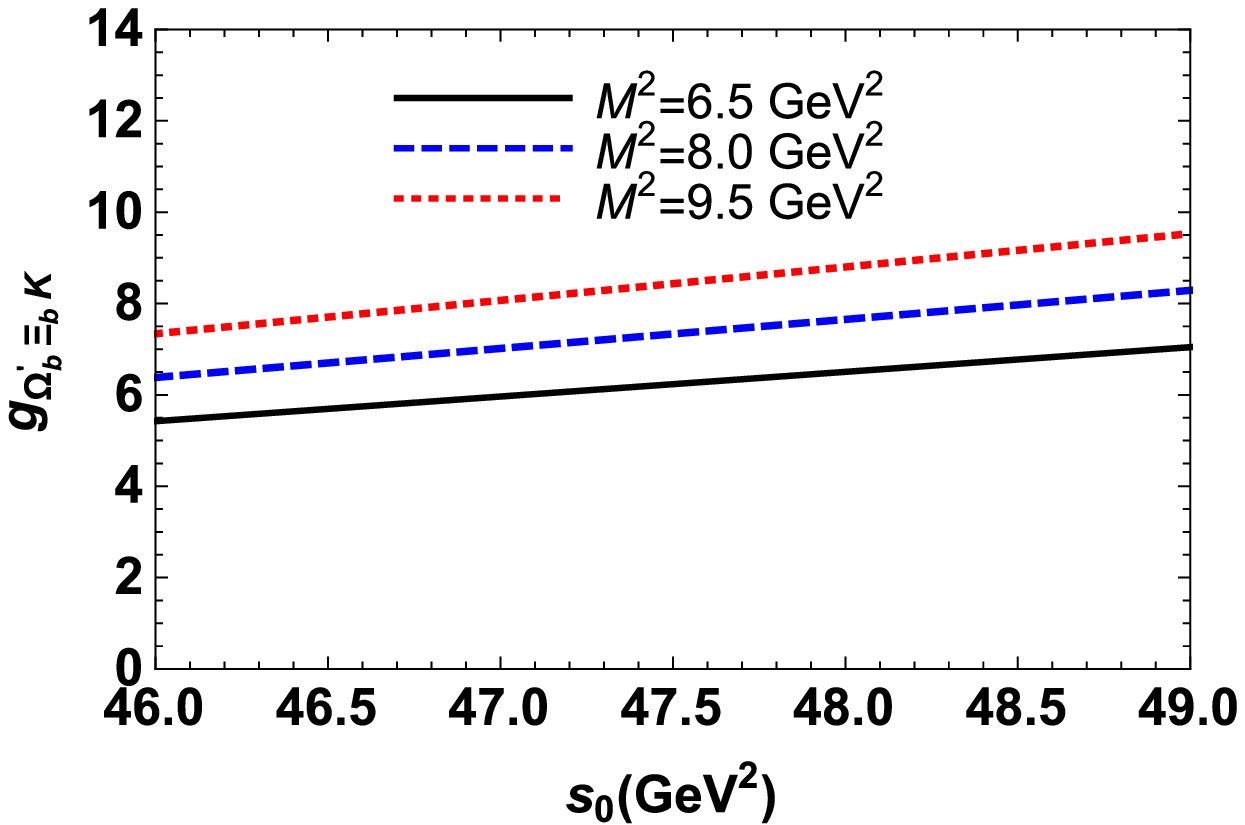}
\end{center}
\caption{ The dependence of the strong coupling   $g_{\Omega_b^{\prime}\Xi_{b}K }$  on the Borel parameter $M^2$ at fixed $s_0$ (left panel), and on the continuum threshold $s_0$ for chosen $M^2$ (right panel).}
\label{fig:Coupl1}
\end{figure}
\begin{figure}[h!]
\begin{center}
\includegraphics[totalheight=6cm,width=8cm]{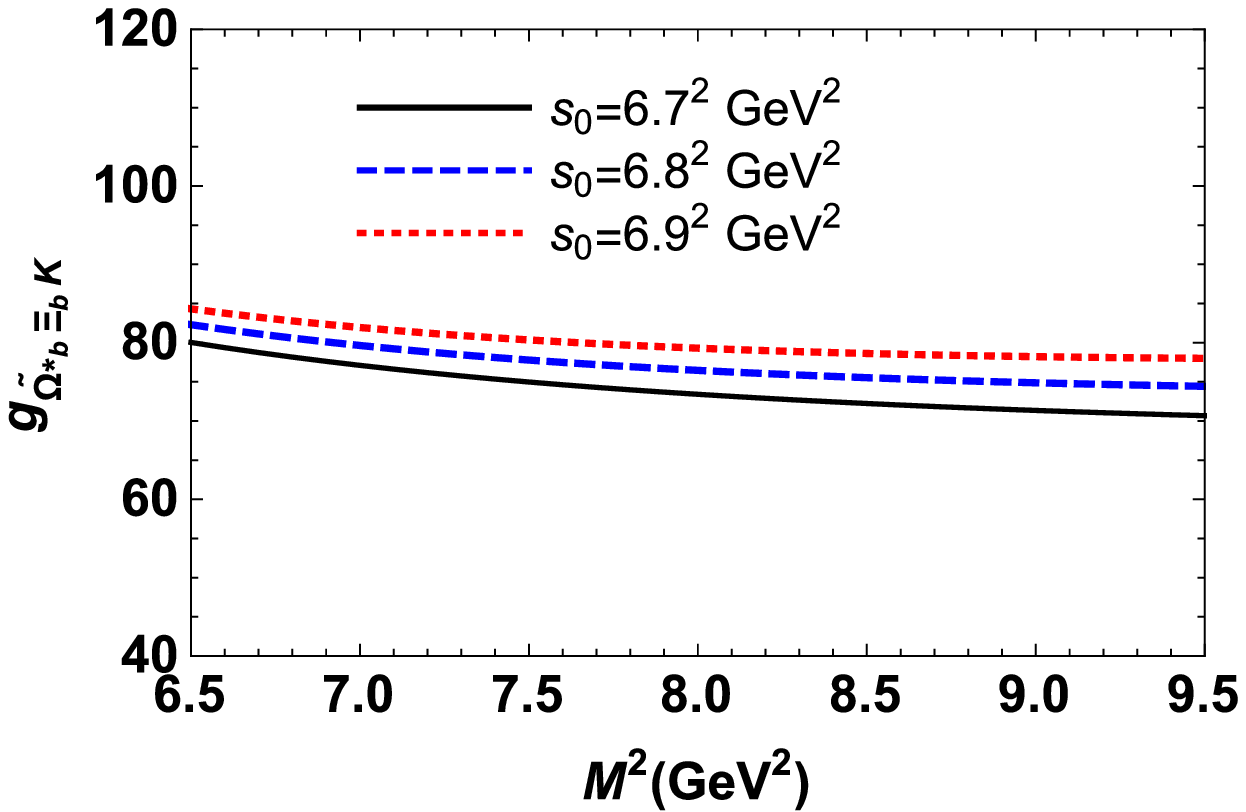}
\includegraphics[totalheight=6cm,width=8cm]{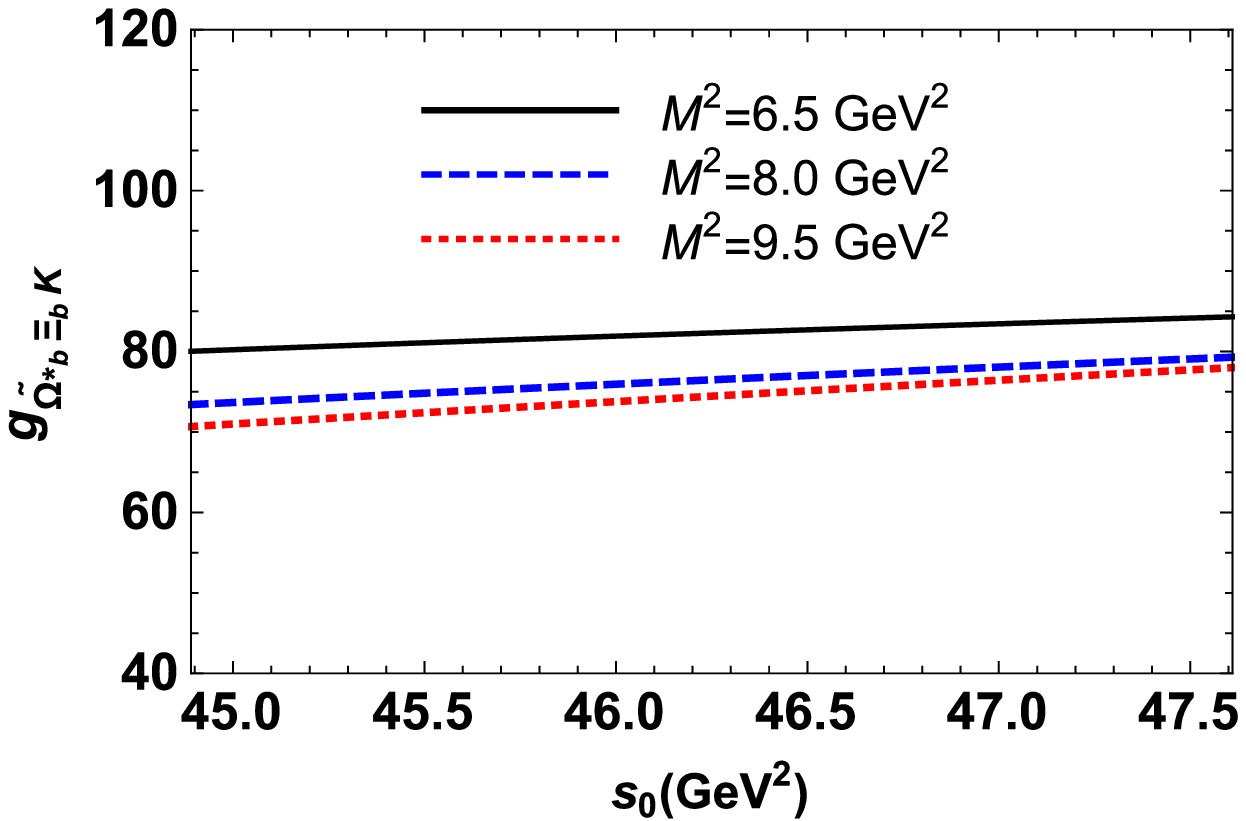}
\end{center}
\caption{ The  strong coupling
$g_{\widetilde\Omega_b^{\star}\Xi_{b}K }$ vs the Borel parameter
$M^2$ (left panel), and vs continuum threshold $s_0$ (right
panel).} \label{fig:Coupl1A}
\end{figure}
\begin{figure}[h!]
\begin{center}
\includegraphics[totalheight=6cm,width=8cm]{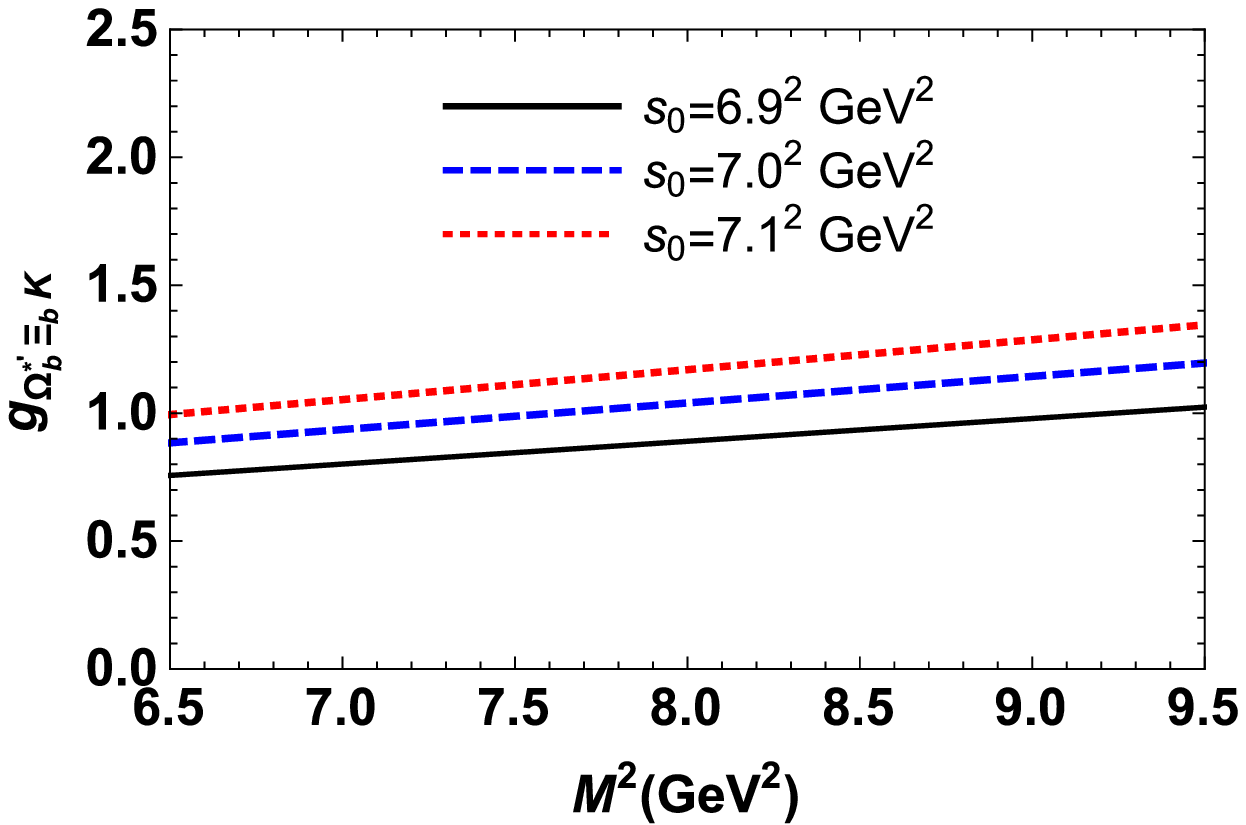}
\includegraphics[totalheight=6cm,width=8cm]{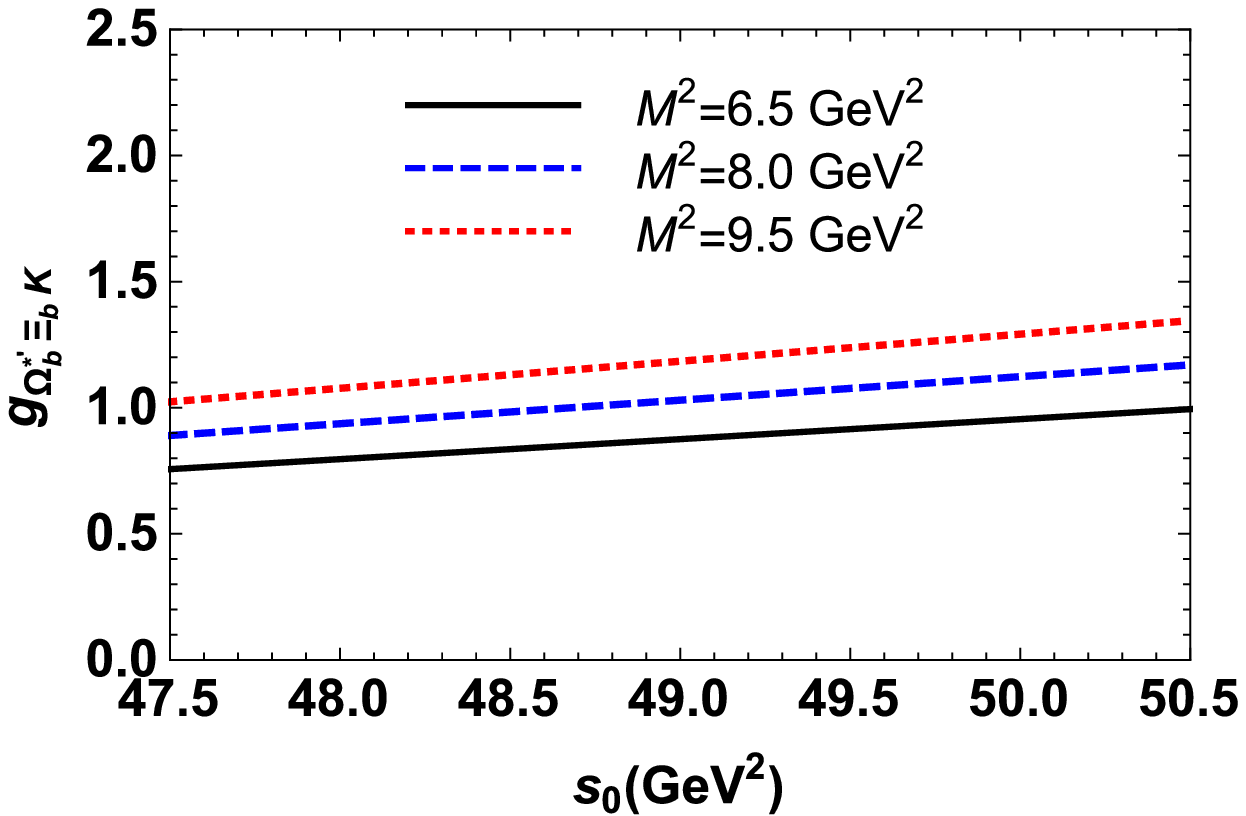}
\end{center}
\caption{ The strong coupling of the radially excited  $\Omega_b^{\star \prime}$ baryon with $\Xi_{b} K$ as a function of the Borel parameter $M^2$ at fixed $s_0$ (left panel), and as a function of the continuum threshold $s_0$ at different $M^2$ (right panel).}
\label{fig:Coupl2}
\end{figure}

\end{widetext}


\section{Concluding remarks}

\label{sec:Conc} 
In the present study we have investigated the decay processes involving the orbitally
and radially excited spin-1/2 and spin-3/2 bottom baryons $\Omega_b$ and $%
\Omega_b^{\star}$, respectively. It is worth noting that the hadronic processes
with heavy baryons and their excitations are interesting from theoretical
point of view, but after discoveries of the LHCb Collaboration they are on
agenda of the experimental collaborations, as well.

In our previous works \cite{Agaev:2017jyt,Agaev:2017lip} we have explained
four of the recently discovered five narrow charmonium-like resonances as the first orbital and
radial excitations of the spin-1/2 and spin-3/2 $\Omega _{c}$ and $\Omega
_{c}^{\star }$ baryons. The masses of their bottom counterparts were already
calculated in Ref.\ \cite{Agaev:2017jyt}. The mass range of the bottom baryons
obtained there indicates that the mass splitting between $(1P,1/2^{-})$ and $%
(1P,3/2^{-})$ baryons, and between $(2S,1/2^{+})$ and $(2S,3/2^{+})$ baryons
is small. At the same time, there is a mass gap between $1P$ and $2S$
states, which may be occupied by "fifth" resonance. In the present work we
have computed the widths of the four $1P$ and $2S$ baryons' decays to $\Xi
_{b}^{0}K^{-}$. The obtained results may be useful for forthcoming
experiments to explore the bottom baryons and measure their spectroscopic and
dynamical parameters.


\appendix*

\section{ The correlation functions and K meson DAs}

\label{sec:App} \renewcommand{\theequation}{\Alph{section}.\arabic{equation}}

In this Appendix we provide explicit expressions for double Borel
transformed form of the invariant amplitude $\Pi _{1}(M^{2})$ for spin-1/2
baryons, as well as the double Borel transformed form of the invariant
amplitude corresponding to the structure $\slashed{q}\slashed{p}\gamma _{\mu
}$ in the correlation function of the spin-3/2 baryons.

\begin{widetext}

For $\Pi _{1}(M^{2})$ we get:
\begin{equation}
\Pi _{1}(M^{2})=\Pi ^{\mathrm{I}}(M^{2})+\Pi ^{\mathrm{\langle s\bar{s}%
\rangle }}(M^{2})+\Pi ^{\mathrm{\langle GG\rangle }}(M^{2})+\Pi ^{\mathrm{%
\langle sG\bar{s}\rangle }}(M^{2})+\Pi ^{\mathrm{\langle s\bar{s}\rangle
\langle GG\rangle }}(M^{2})+\Pi ^{\mathrm{\langle sG\bar{s}\rangle \langle
GG\rangle }}(M^{2}),
\end{equation}%
\begin{eqnarray}
\Pi ^{\mathrm{I}}(M^{2}) &=&\frac{1}{96\sqrt{2}\pi ^{2}}\int_{m_{b}^{2}}^{%
\infty }dse^{\frac{m_{K}^{2}-4s}{4M^{2}}}\frac{m_{b}}{s^{3}}\Bigg\lbrace%
\sqrt{3}m_{b}^{2}M^{2}\Bigg[3f_{K}m_{K}^{2}(1-\beta ^{2})s\mathbb{A}%
(u_{0})-12f_{K}M^{2}(1-\beta )\Big[(1+\beta )(s-m_{b}^{2})  \notag \\
&+&\beta m_{b}m_{s}\Big]\phi _{K}(u_{0})-4\mu _{K}(\widetilde{\mu }%
_{K}^{2}-1)\Big[(\beta -1)(2\beta +1)M^{2}m_{b}+2(1+\beta +\beta ^{2})sm_{s}%
\Big]\phi _{\sigma }(u_{0})\Bigg]+f_{K}m_{K}^{2}(\beta -1)  \notag \\
&+&\Bigg(M^{2}\Big[(\beta -1)s^{2}+2\beta sm_{b}m_{s}+(3+\beta
)sm_{b}^{2}-(\beta -1)m_{b}^{3}m_{s}\Big]+(\beta -1)m_{b}^{3}m_{s}(s+2M^{2})%
\mathrm{Ln}[\Psi ]\Bigg)  \notag \\
&\times &I_{1}\Big(\mathcal{A}_{\parallel }(\alpha ),1\Big)%
+2f_{K}m_{K}^{2}(\beta -1)(1+3\beta )\Bigg(M^{2}\Big[%
m_{b}^{3}m_{s}+sm_{b}(2m_{b}-m_{s})-s^{2}\Big]+m_{b}^{3}m_{s}(2M^{2}+s)%
\mathrm{Ln}[\Psi ]\Bigg)  \notag \\
&\times &I_{1}\Big(\mathcal{A}_{\perp }(\alpha ),1\Big)+2f_{K}m_{K}m_{b}(%
\beta -1)\Bigg(M^{2}\Big[sm_{s}+sm_{b}(1+\beta )+m_{b}^{2}m_{s}(1+2\beta )%
\Big]+(1+2\beta )(s+2M^{2})m_{b}^{2}  \notag \\
&\times &m_{s}\mathrm{Ln}[\Psi ]\Bigg)I_{1}\Big(\mathcal{V}_{\parallel
}(\alpha ),1\Big)+2f_{K}m_{K}^{2}(\beta -1)\Bigg(M^{2}\Big[2\beta
mb^{3}m_{s}+2(1+2\beta ^{2})s+(3+\beta )sm_{b}m_{s}-(5+3\beta )sm_{b}^{2}%
\Big]  \notag \\
&+&2\beta (s+2M^{2})m_{b}^{3}m_{s}\mathrm{Ln}[\Psi ]\Bigg)I_{1}\Big(\mathcal{%
V}_{\perp }(\alpha ),1\Big)+4(\beta -1)\mu _{K}M^{2}m_{K}^{2}m_{b}\Big[%
4mb^{2}(1+2\beta )-3s(1+\beta )\Big]I_{1}\Big(\mathcal{T}(\alpha ),1\Big)
\notag \\
&+&4(\beta -1)f_{K}m_{K}^{2}M^{2}m_{b}s[(1+\beta )m_{b}+\beta m_{s}]I_{1}%
\Big(\mathcal{A}_{\parallel }(\alpha ),v\Big)+4(1-\beta )(3+\beta
)sf_{K}m_{K}^{2}M^{2}m_{b}(m_{b}-m_{s})  \notag \\
&\times &I_{1}\Big(\mathcal{V}_{\perp }(\alpha ),v\Big)+18(\beta -1)\mu
_{K}M^{2}m_{K}^{2}m_{b}[m_{b}^{2}(1+\beta )-s]u_{0}I_{1}\Big(\mathcal{T}%
(\alpha ),v\Big)+(\beta -1)\mu _{K}M^{4}m_{b}  \notag \\
&\times &[4m_{b}^{2}(1+2\beta )-3s(1+\beta )]I_{2}\Big(\mathcal{T}(\alpha ),1%
\Big)-4(1-\beta )\mu _{K}M^{4}m_{b}[m_{b}^{2}(1+\beta )-s]I_{2}\Big(\mathcal{%
T}(\alpha ),v\Big)\Bigg\rbrace  \notag \\
&+&\dfrac{(1-\beta )}{32\sqrt{6}\pi ^{2}}e^{\dfrac{m_{K}^{2}-4m_{b}^{2}}{%
4M^{2}}}f_{K}m_{K}^{2}M^{2}m_{s}\Bigg\lbrace t\mathbb{A}(u_{0})+\gamma _{E}%
\Bigg[(1-\beta )I_{1}\Big(\mathcal{A}_{\parallel }(\alpha ),1\Big)%
+2(1+3\beta )I_{1}\Big(\mathcal{A}_{\perp }(\alpha ),1\Big)  \notag \\
&+&2(1+2\beta )I_{1}\Big(\mathcal{V}_{\parallel }(\alpha ),1\Big)-4\beta
I_{1}\Big(\mathcal{V}_{\perp }(\alpha ),1\Big)\Bigg]\Bigg\rbrace,
\label{PiQCD}
\end{eqnarray}%
\begin{eqnarray}
\Pi ^{\mathrm{\langle s\bar{s}\rangle }}(M^{2}) &=&\dfrac{\langle s\bar{s}%
\rangle }{144\sqrt{6}M^{4}}e^{\dfrac{m_{K}^{2}-4m_{b}^{2}}{4M^{2}}}%
\Bigg\lbrace3f_{K}m_{K}^{2}(\beta -1)\Big[m_{b}^{3}m_{s}(1+\beta )-2\beta
M^{2}(m_{b}^{2}-M^{2})\Big]\mathbb{A}(u_{0})+12M^{2}(\beta -1)f_{K}  \notag
\\
&\times &M^{2}\Big(m_{b}m_{s}(1+\beta )-2\beta M^{2}\Big)\phi
_{K}(u_{0})-4M^{2}\mu _{K}(1-\tilde{\mu}_{K}^{2})\Bigg(4M^{2}m_{b}(1+\beta
+\beta ^{2})+m_{b}^{2}m_{s}(1+\beta -2\beta ^{2})  \notag \\
&-&M^{2}m_{s}(1+\beta -2\beta ^{2})\Bigg)\phi _{\sigma }(u_{0})+3M^{2}(\beta
-1)f_{K}m_{K}^{2}\Bigg[\Big(m_{b}m_{s}(\beta -1)-4\beta M^{2}\Big)I_{1}\Big(%
\mathcal{A}_{\parallel }(\alpha ),1\Big)+8\beta M^{2}  \notag \\
&\times &I_{1}\Big(\mathcal{A}_{\parallel }(\alpha ),v\Big)-2(1+3\beta
)(m_{b}m_{s}-2M^{2})I_{1}\Big(\mathcal{A}_{\perp }(\alpha ),1\Big)-4\Big(%
m_{b}m_{s}(1+\beta )-M^{2}(3+\beta )\Big)I_{1}\Big(\mathcal{V}_{\perp
}(\alpha ),1\Big)  \notag \\
&-&8(3+\beta )M^{2}I_{1}\Big(\mathcal{V}_{\perp }(\alpha ),v\Big)-4M^{2}I_{1}%
\Big(\mathcal{V}_{\parallel }(\alpha ),1\Big)\Bigg]-3M^{2}(\beta -1)\mu _{K}%
\Bigg[16m_{s}m_{K}^{2}u_{0}I_{1}\Big(\mathcal{T}(\alpha ),v\Big)  \notag \\
&-&12m_{K}^{2}m_{s}(1+\beta )u_{0}I_{1}\Big(\mathcal{T}(\alpha ),1\Big)%
-3M^{2}m_{s}(1+\beta )I_{2}\Big(\mathcal{T}(\alpha ),1\Big)+4M^{2}m_{s}I_{2}%
\Big(\mathcal{T}(\alpha ),v\Big)\Bigg]\Bigg\rbrace,  \label{PiQCDssbar}
\end{eqnarray}%
\begin{eqnarray}
\Pi ^{\mathrm{\langle GG\rangle }}(M^{2}) &=&\dfrac{\langle GG\rangle }{6912%
\sqrt{6}\pi ^{2}M^{8}}\Bigg\lbrace\int_{m_{b}^{2}}^{\infty }dse^{\frac{%
m_{K}^{2}-4s}{4M^{2}}}\Bigg\lbrace\dfrac{3m_{b}}{s^{3}}\Bigg[3\beta (\beta
-1)f_{K}m_{K}^{2}m_{b}^{3}m_{s}\Big(M^{2}(2M^{4}+3M^{2}s+3s^{2})+s^{3}%
\mathrm{Ln}[\Psi ]\Big)\mathbb{A}(u_{0})  \notag \\
&+&12(\beta -1)M^{4}f_{K}\Big[M^{6}s+\beta M^{2}\Big(%
M^{4}s-m_{b}^{3}m_{s}(3M^{2}+2s)\Big)-\beta
(2M^{4}+2M^{2}s+s^{2})m_{b}^{3}m_{s}\mathrm{Ln}[\Psi ]\Big]\phi _{K}(u_{0})
\notag \\
&-&8(1+\beta +\beta ^{2})\mu _{K}(1-\tilde{\mu}_{K}^{2})M^{2}m_{b}^{2}m_{s}s%
\Big[M^{2}(M^{2}+2s)+s^{2}\mathrm{Ln}[\Psi ]\Big]\phi _{\sigma }(u_{0})\Bigg]%
-\dfrac{3M^{2}}{s^{2}}(\beta -1)f_{K}m_{K}^{2}m_{b}^{2}m_{s}  \notag \\
&\times &\Big(M^{2}(M^{2}+2s)+s^{2}\mathrm{Ln}[\Psi ]\Big)\Bigg[(1+5\beta
)I_{1}\Big(\mathcal{A}_{\parallel }(\alpha ),1\Big)-4I_{1}\Big(\mathcal{A}%
_{\perp }(\alpha ),1\Big)-12\beta \Big(I_{1}\Big(\mathcal{A}_{\parallel
}(\alpha ),v\Big)  \notag \\
&+&I_{1}\Big(\mathcal{A}_{\perp }(\alpha ),1\Big)\Big)+4(2+\beta )I_{1}\Big(%
\mathcal{V}_{\parallel }(\alpha ),1\Big)-2(9+5\beta )I_{1}\Big(\mathcal{V}%
_{\perp }(\alpha ),1\Big)+12(3+\beta )I_{1}\Big(\mathcal{V}_{\perp }(\alpha
),v\Big)\Bigg]\Bigg\rbrace  \notag \\
&+&e^{\frac{m_{K}^{2}-4m_{b}^{2}}{4M^{2}}}\Bigg\lbrace\dfrac{M^{2}}{m_{b}}%
\Bigg[3(1-\beta )f_{K}m_{K}^{2}\Big[\beta (3\gamma
_{E}-2)m_{b}^{5}m_{s}-\beta M^{2}m_{b}^{3}m_{s}-(1+\beta
)M^{4}m_{b}^{2}+2(1+\beta )M^{6}\Big]\mathbb{A}(u_{0})  \notag \\
&+&12(1-\beta )f_{K}M^{4}\Big[\beta (2-3\gamma _{E})m_{b}^{3}m_{s}+M^{2}\Big(%
M^{2}+\beta (3(1-\gamma _{E})m_{b}m_{s}+M^{2})\Big)\Big]\phi
_{K}(u_{0})+4\mu _{K}(1-\tilde{\mu}_{K}^{2})M^{2}  \notag \\
&\times &\Big[(1+\beta -2\beta ^{2})M^{4}m_{b}+2(1+\beta +\beta ^{2})\Big(%
m_{b}^{2}m_{s}((3\gamma _{E}-2)m_{b}^{2}-M^{2})+2M^{4}m_{s}\Big)\Big]\phi
_{\sigma }(u_{0})\Bigg]  \notag \\
&+&\dfrac{3M^{4}}{m_{b}}(\beta -1)f_{K}m_{K}^{2}\Bigg[\Big((1+5\beta )\gamma
_{E}m_{b}^{3}m_{s}-4\beta m_{b}^{3}m_{s}-2(1+\beta )M^{2}m_{b}^{2}+4(1+\beta
)M^{4}\Big)I_{1}\Big(\mathcal{A}_{\parallel }(\alpha ),1\Big)  \notag \\
&-&2(1+3\beta )\Big(2(\gamma _{E}-1)m_{b}^{3}m_{s}+M^{2}m_{b}^{2}-2M^{4}\Big)%
I_{1}\Big(\mathcal{A}_{\perp }(\alpha ),1\Big)+2\Big((1+\beta
)M^{2}(2M^{2}-m_{b}^{2})+2m_{b}^{3}m_{s}  \notag \\
&\times &(\gamma _{E}(2+\beta )-1)\Big)I_{1}\Big(\mathcal{V}_{\parallel
}(\alpha ),1\Big)-2\Big((3+\beta
)M^{2}(m_{b}^{2}-2M^{2})+m_{b}^{3}m_{s}(\gamma _{E}(9+5\beta )-3\beta -6)%
\Big)  \notag \\
&\times &I_{1}\Big(\mathcal{V}_{\perp }(\alpha ),1\Big)+4(3+\beta )\Big(%
(3\gamma _{E}-2)m_{b}^{3}m_{s}+M^{2}(m_{b}^{2}-2M^{2})\Big)I_{1}\Big(%
\mathcal{V}_{\perp }(\alpha ),v\Big)+4\Big(\beta (2-3\gamma
_{E})m_{b}^{3}m_{s}  \notag \\
&+&M^{2}(1+\beta )(m_{b}^{2}-2M^{2})\Big)I_{1}\Big(\mathcal{A}_{\parallel
}(\alpha ),v\Big)\Bigg]+\dfrac{3M^{4}}{m_{b}}(\beta -1)\mu _{K}\Bigg[%
4(1+5\beta )m_{K}^{2}M^{2}m_{b}u_{0}I_{1}\Big(\mathcal{T}(\alpha ),1\Big)
\notag \\
&-&16\beta m_{K}^{2}M^{2}m_{b}u_{0}I_{1}\Big(\mathcal{T}(\alpha ),v\Big)%
+(1+5\beta )M^{4}m_{b}I_{2}\Big(\mathcal{T}(\alpha ),1\Big)-4\beta
M^{4}m_{b}I_{2}\Big(\mathcal{T}(\alpha ),v\Big)\Bigg]\Bigg\rbrace\Bigg\rbrace%
,  \label{PiQCDGG}
\end{eqnarray}%
\begin{eqnarray}
\Pi ^{\mathrm{\langle sG\bar{s}\rangle }}(M^{2}) &=&\dfrac{m_{0}^{2}\langle s%
\bar{s}\rangle }{3456\sqrt{6}M^{8}}e^{\dfrac{m_{K}^{2}-4m_{b}^{2}}{4M^{2}}}%
\Bigg\lbrace3f_{K}m_{K}^{2}m_{b}(\beta -1)\Big[4m_{b}^{2}m_{s}(1+\beta
)(m_{b}^{2}-3M^{2})-12\beta m_{b}^{3}M^{2}+4M^{4}m_{s}  \notag \\
&\times &(1+\beta )+M^{4}m_{b}\Big(t-11+2(7+\beta )v\Big)\Big]\mathbb{A}%
(u_{0})+12(\beta -1)f_{K}M^{4}\Big(4m_{b}m_{s}(1+\beta )(m_{b}^{2}-M^{2})
\notag \\
&+&M^{4}(2v(7+\beta )-11(1+\beta ))\Big)\phi _{K}(u_{0})+8M^{2}\mu _{K}(%
\tilde{\mu}_{K}^{2})m_{b}\Big[2(2\beta +1)(\beta -1)m_{b}^{3}m_{s}+(4+\beta
-5\beta ^{2})  \notag \\
&\times &M^{2}m_{b}m_{s}-12(1+\beta +\beta
^{2})M^{2}m_{b}^{2}+3M^{4}(3+2\beta +3\beta ^{2})\Big]\phi _{\sigma
}+12M^{2}(1-\beta )f_{K}m_{K}^{2}\Bigg[\Big(m_{s}(\beta -1)  \notag \\
&\times &(m_{b}^{2}-M^{2})-6\beta M^{2}m_{b}\Big)I_{1}\Big(\mathcal{A}%
_{\parallel }(\alpha ),1\Big)+12\beta M^{2}m_{b}I_{1}\Big(\mathcal{A}%
_{\parallel }(\alpha ),v\Big)+2(1+3\beta )\Big(M^{2}(3m_{b}+m_{s})  \notag \\
&-&m_{b}^{2}m_{s}\Big)I_{1}\Big(\mathcal{A}_{\perp }(\alpha ),1\Big)+2\Big(%
2(1+\beta )m_{s}(M^{2}-m_{b}^{2})+3(3+\beta )M^{2}m_{b}\Big)I_{1}\Big(%
\mathcal{V}_{\perp }(\alpha ),1\Big)-12(3+\beta )  \notag \\
&\times &M^{2}m_{b}I_{1}\Big(\mathcal{V}_{\perp }(\alpha ),v\Big )%
-6M^{2}m_{b}I_{1}\Big(\mathcal{V}_{\parallel }(\alpha ),1\Big )\Bigg]+12\mu
_{K}M^{2}(1-\beta )\Bigg[12(1+\beta )m_{K}^{2}m_{b}m_{s}u_{0}I_{1}\Big(%
\mathcal{T}(\alpha ),1\Big )  \notag \\
&-&16m_{K}^{2}m_{b}m_{s}u_{0}I_{1}\Big(\mathcal{T}(\alpha ),v\Big )%
+3(1+\beta )M^{2}m_{b}m_{s}I_{2}\Big(\mathcal{T}(\alpha ),1\Big )%
-4M^{2}m_{b}m_{s}I_{2}\Big(\mathcal{T}(\alpha ),v\Big )\Bigg]\Bigg\rbrace,
\label{PiQCDm02ssbar}
\end{eqnarray}%
\begin{eqnarray}
\Pi ^{\mathrm{\langle s\bar{s}\rangle \langle GG\rangle }}(M^{2}) &=&\dfrac{%
\langle s\bar{s}\rangle \langle GG\rangle }{10368\sqrt{6}M^{10}}e^{\frac{%
m_{K}^{2}-4m_{b}^{2}}{4M^{2}}}m_{b}\Bigg\lbrace3f_{K}m_{K}^{2}(\beta -1)\Big[%
2\beta M^{2}m_{b}(2M^{2}-m_{b}^{2})+(1+\beta
)m_{b}^{2}m_{s}(m_{b}^{2}-6M^{2})  \notag \\
&+&6(1+\beta )M^{4}m_{s}\Big]\mathbb{A}(u_{0})+4M^{2}\Big[3(\beta
-1)f_{K}M^{2}\Big(2\beta M^{2}m_{b}+(1+\beta )m_{s}(m_{b}^{2}+3M^{2})\Big)%
\phi _{K}(u_{0})  \notag \\
&+&\mu _{K}(\tilde{\mu}_{K}^{2}-1)\Big((\beta -1)(1+2\beta
)m_{b}^{3}m_{s}+2(1+\beta -2\beta ^{2})M^{2}m_{b}m_{s}  \notag \\
&-&4(1+\beta +\beta ^{2})M^{2}(m_{b}^{2}-3M^{2})\Big)\phi _{\sigma }(u_{0})%
\Big]\Bigg\rbrace,  \label{PiQCDssGG}
\end{eqnarray}%
\begin{eqnarray}
\Pi ^{\mathrm{\langle sG\bar{s}\rangle \langle GG\rangle }}(M^{2}) &=&\dfrac{%
m_{0}^{2}\langle s\bar{s}\rangle \langle GG\rangle }{62208\sqrt{6}M^{14}}e^{%
\dfrac{m_{K}^{2}-4m_{b}^{2}}{4M^{2}}}m_{b}\Bigg\lbrace3f_{K}m_{K}^{2}(1-%
\beta )\Big[3\beta M^{2}m_{b}(6M^{2}m_{b}^{2}-m_{b}^{4}-6M^{4})-(1+\beta
)M^{2}  \notag \\
&\times &m_{s}(11m_{b}^{4}-30M^{2}m_{b}^{2}+18M^{4})\Big]\mathbb{A}%
(u_{0})+12f_{K}M^{4}(\beta -1)\Big[3\beta M^{2}m_{b}(2M^{2}-m_{b}^{2})
\notag \\
&+&(1+\beta )m_{s}(m_{b}^{4}-6M^{2}m_{b}^{2}+6M^{4})\Big]\phi
_{K}(u_{0})+\mu _{K}(\tilde{\mu}_{K}^{2}-1)\Big(%
m_{b}^{4}-6M^{2}(m_{b}^{2}-M^{2})\Big)  \notag \\
&\times &\Big[(1+\beta -2\beta ^{2})m_{b}m_{s}+6(1+\beta +\beta ^{2})M^{2}%
\Big]\phi _{\sigma }(u_{0})\Bigg\rbrace,  \label{PiQCDm02ssGG}
\end{eqnarray}

For the structure $\slashed{q}\slashed{p}\gamma _{\mu }$ in spin-$3/2$
baryons' correlation function we find:
\begin{eqnarray}
\widetilde{\Pi }^{\mathrm{I}}(M^{2}) &=&\frac{m_{b}}{96\sqrt{2}\pi ^{2}}%
\int_{m_{b}^{2}}^{\infty }dse^{\frac{m_{K}^{2}-4s}{4M^{2}}}\dfrac{M^{2}}{%
s^{3}}\Bigg\{3f_{K}m_{b}^{2}(1+\beta )\Big[4M^{2}(s-m_{b}^{2})\phi
_{K}(u_{0})-sm_{K}^{2}\mathbb{A}(u_{0})\Big]-4\beta m_{b}^{2}(\widetilde{\mu
}_{K}^{2}-1)\mu _{K}  \notag \\
&&\times M^{2}m_{b}\phi _{\sigma }(u_{0})+4m_{b}^{2}m_{s}\Big[3\beta
f_{K}M^{2}m_{b}\phi _{K}(u_{0})+(1-\beta )\mu _{K}(\tilde{\mu}%
_{K}^{2}-1)s\phi _{\sigma }(u_{0})\Big]+I_{1}\Big(\mathcal{A}_{\parallel
}(\alpha ),1\Big)f_{K}m_{K}^{2}s  \notag \\
&&\times \Big(s(1-\beta )+2m_{b}^{2}(1+2\beta )\Big)+4I_{1}\Big(\mathcal{A}%
_{\perp }(\alpha ),1\Big)f_{K}m_{K}^{2}m_{b}^{2}s(1+2\beta )+3I_{1}\Big(%
\mathcal{V}_{\parallel }(\alpha ),1\Big)f_{K}m_{K}^{2}m_{b}^{2}s(1+\beta )
\notag \\
&&+2I_{1}\Big(\mathcal{V}_{\perp }(\alpha ),1\Big)f_{K}m_{K}^{2}s\Big[%
s(1-\beta )+3m_{b}^{2}(1+\beta )\Big]-4I_{1}\Big(\mathcal{T}(\alpha ),1\Big)%
\mu _{K}m_{K}^{2}m_{b}u_{0}\Big[2s(1+2\beta )-m_{b}^{2}(1-\beta )\Big]
\notag \\
&&-6I_{1}\Big(\mathcal{A}_{\parallel }(\alpha ),v\Big)%
f_{K}m_{K}^{2}m_{b}^{2}s(1+\beta )-8I_{1}\Big(\mathcal{V}_{\perp }(\alpha ),v%
\Big)f_{K}m_{K}^{2}m_{b}^{2}s(2+\beta )+8I_{1}\Big(\mathcal{T}(\alpha ),v%
\Big)\mu _{K}m_{K}^{2}m_{b}u_{0}  \notag \\
&&\times \Big[s(2+\beta )-m_{b}^{2}(1-\beta )\Big]+I_{2}\Big(\mathcal{T}%
(\alpha ),1\Big)\mu _{K}M^{2}m_{b}\Big[m_{b}^{2}(1-\beta )-2s(1+2\beta )\Big]%
+2I_{2}\Big(\mathcal{T}(\alpha ),v\Big)\mu _{K}M^{2}m_{b}  \notag \\
&&\times \Big[m_{b}^{2}(1-\beta )-2s(1+2\beta )\Big]%
-2f_{K}m_{K}^{2}m_{b}m_{s}\Bigg[2\Big[\Big(m_{b}^{2}(1+2\beta )-2(2+\beta )%
\Big)+(1+2\beta )m_{b}^{2}(2+\dfrac{s}{M^{2}})\mathrm{Ln}[\Psi ]\Big]  \notag
\\
&&\times I_{1}\Big(\mathcal{V}_{\perp }(\alpha ),1\Big)-3\Big[(s+\beta
m_{b}^{2})+\beta m_{b}^{2}(2+\dfrac{s}{M^{2}})\mathrm{Ln}[\Psi ]\Big]I_{1}%
\Big(\mathcal{V}_{\parallel }(\alpha ),1\Big)-\Big[\Big((1+2\beta
)m_{b}^{2}-3\beta s\Big)+(1+2\beta )  \notag \\
&&\times m_{b}^{2}(2+\dfrac{s}{M^{2}})\mathrm{Ln}[\Psi ]\Big]I_{1}\Big(%
\mathcal{A}_{\parallel }(\alpha ),1\Big)+2\Big[(1+\beta )s+3\beta
m_{b}^{2}+3\beta m_{b}^{2}(2+\frac{s}{M^{2}})\mathrm{Ln}[\Psi ]\Big]I_{1}%
\Big(\mathcal{A}_{\perp }(\alpha ),1\Big)  \notag \\
&&-6\beta sI_{1}\Big(\mathcal{A}_{\parallel }(\alpha ),v\Big)-4(2+\beta
)sI_{1}\Big(\mathcal{V}_{\perp }(\alpha ),v\Big)\Bigg]\Bigg\}  \notag \\
&&+\dfrac{m_{s}}{96\sqrt{6}\pi ^{2}}e^{\dfrac{m_{K}^{2}-4m_{b}^{2}}{4M^{2}}%
}f_{K}m_{K}^{2}M^{2}\Bigg\lbrace2\gamma _{E}\Big[(1+2\beta )I_{1}\Big(%
\mathcal{A}_{\parallel }(\alpha ),1\Big)-3\beta \Big(2I_{1}\Big(\mathcal{A}%
_{\perp }(\alpha ),1\Big)+I_{1}\Big(\mathcal{V}_{\parallel }(\alpha ),1\Big)%
\Big)  \notag \\
&&+2(1+2\beta )I_{1}\Big(\mathcal{V}_{\perp }(\alpha ),1\Big)\Big]-3\beta
\mathbb{A}(u_{0})\Bigg\rbrace,  \label{eq:PiI}
\end{eqnarray}%
\begin{eqnarray}
\widetilde{\Pi }^{\mathrm{\langle s\bar{s}\rangle }}(M^{2}) &=&\frac{\langle
s\bar{s}\rangle }{72\sqrt{2}M^{2}}e^{\frac{m_{K}^{2}-4m_{b}^{2}}{4M^{2}}}%
\Bigg\{3\beta f_{K}m_{K}^{2}(M^{2}+m_{b}^{2})\mathbb{A}(u_{0})-4M^{2}\Big[%
3\beta f_{K}M^{2}\phi _{K}(u_{0})+\mu _{K}m_{b}(\widetilde{\mu }%
_{K}^{2}-1)\phi _{\sigma }(u_{0})\Big]  \notag \\
&&-2M^{2}f_{K}m_{K}^{2}\Big[3\beta I_{1}\Big(\mathcal{A}_{\parallel }(\alpha
),1\Big)-2(1+2\beta )I_{1}\Big(\mathcal{A}_{\perp }(\alpha ),1\Big)-3I_{1}%
\Big(\mathcal{V}_{\parallel }(\alpha ),1\Big)-2(2+\beta )I_{1}\Big(\mathcal{V%
}_{\perp }(\alpha ),1\Big)  \notag \\
&&+6\beta I_{1}\Big(\mathcal{A}_{\parallel }(\alpha ),v\Big)+4(2+\beta )I_{1}%
\Big(\mathcal{V}_{\perp }(\alpha ),v\Big)\Big]-\dfrac{m_{s}}{2}\Bigg[%
3(1+\beta )f_{K}m_{K}^{2}m_{b}^{3}\mathbb{A}(u_{0})-4M^{2}\Big[3(1+\beta
)f_{K}M^{2}  \notag \\
&&\times m_{b}\phi _{K}(u_{0})-\beta \mu _{K}(\tilde{\mu}%
_{K}^{2}-1)(M^{2}+m_{b}^{2})\phi _{\sigma }(u_{0})\Big]\Bigg]%
+f_{K}m_{K}^{2}m_{s}(1-\beta )m_{b}\Big(I_{1}\Big(\mathcal{A}_{\parallel
}(\alpha ),1\Big)+2I_{1}\Big(\mathcal{V}_{\perp }(\alpha ),1\Big)\Big)
\notag \\
&&-8\mu _{K}m_{K}^{2}m_{s}u_{0}\Big[(1+2\beta )I_{1}\Big(\mathcal{T}(\alpha
),1\Big)-(2+\beta )I_{1}\Big(\mathcal{T}(\alpha ),v\Big)\Big]-2\mu
_{K}M^{2}m_{s}\Big[(1+2\beta )I_{2}\Big(\mathcal{T}(\alpha ),1\Big)  \notag
\\
&&-(2+\beta )I_{2}\Big(\mathcal{T}(\alpha ),v\Big)\Big]\Bigg\},
\label{eq:Piqqbar}
\end{eqnarray}%
\begin{eqnarray}
\widetilde{\Pi }^{\mathrm{\langle GG\rangle }}(M^{2}) &=&\dfrac{\langle
GG\rangle }{192\sqrt{2}\pi ^{2}}\Bigg\{\int_{m_{b}^{2}}^{\infty }dse^{\frac{%
m_{K}^{2}-4s}{4M^{2}}}\Bigg[\dfrac{M^{2}f_{K}m_{b}(1+\beta )}{s^{2}}\phi
_{K}(u_{0})+\dfrac{m_{b}^{3}m_{s}}{12M^{8}s^{3}}\Bigg(3\beta f_{K}m_{b}\Big(%
m_{K}^{2}(M^{2}(2M^{2}+3M^{2}s+3s^{2})  \notag \\
&+&s^{3}\mathrm{Ln}[\Psi ])\mathbb{A}%
(u_{0})-4M^{4}(M^{2}(3M^{2}+2s)+(2M^{4}+2M^{2}s+s^{2})\mathrm{Ln}[\Psi
])\phi _{K}(u_{0})+4(\beta -1)\mu _{K}(\tilde{\mu}_{K}^{2}-1)M^{2}s  \notag
\\
&\times &(M^{2}(2s+M^{2})+s^{2}\mathrm{Ln}[\Psi ])\phi _{\sigma }(u_{0})\Big)%
\Bigg)-\dfrac{f_{K}m_{K}^{2}m_{b}^{2}m_{s}}{18M^{6}s^{2}}(M^{4}+2M^{2}s+s^{2}%
\mathrm{Ln}[\Psi ])\Bigg((7\beta -1)I_{1}\Big(\mathcal{A}_{\parallel
}(\alpha ),1\Big)  \notag \\
&+&(9+3\beta )I_{1}\Big(\mathcal{V}_{\parallel }(\alpha ),1\Big)+2(\beta
+5)I_{1}\Big(\mathcal{V}_{\perp }(\alpha ),1\Big)+6(1+3\beta )I_{1}\Big(%
\mathcal{A}_{\perp }(\alpha ),1\Big)-18\beta I_{1}\Big(\mathcal{A}%
_{\parallel }(\alpha ),v\Big)  \notag \\
&-&12(2+\beta )I_{1}\Big(\mathcal{A}_{\perp }(\alpha ),v\Big)\Bigg)\Bigg]+%
\dfrac{1}{36M^{2}m_{b}}e^{\frac{m_{K}^{2}-4m_{b}^{2}}{4M^{2}}}\Bigg[%
3f_{K}(1+\beta \mathbb{A}(u_{0}))\Big(m_{K}^{2}(m_{b}^{2}-2M^{2})  \notag \\
&-&4M^{4}\phi _{K}(u_{0})\Big)+4\beta \mu _{K}(\widetilde{\mu }%
_{K}^{2}-1)M^{2}m_{b}\phi _{\sigma }(u_{0})+\dfrac{m_{s}}{M^{4}}\Bigg[3\beta
f_{K}m_{b}\Bigg(m_{K}^{2}m_{b}^{2}(M^{2}+(2-3\gamma _{E})m_{b}^{2})\mathbb{A}%
(u_{0})  \notag \\
&+&4M^{4}((3\gamma _{E}-2)m_{b}^{2}+3(\gamma _{E}-1)M^{2})\phi _{K}(u_{0})%
\Bigg)+4(1-\beta )\mu _{K}(\tilde{\mu}_{K}^{2}-1)M^{2}\Big((3\gamma
_{E}-2)m_{b}^{4}-M^{2}m_{b}^{2}+2M^{4}\Big)  \notag \\
&\times &\phi _{\sigma }(u_{0})\Bigg]+6I_{1}\Big(\mathcal{A}_{\parallel
}(\alpha ),1\Big)f_{K}m_{K}^{2}(2M^{2}-m_{b}^{2})(1+\beta )+8I_{1}\Big(%
\mathcal{A}_{\perp }(\alpha ),1\Big)f_{K}m_{K}^{2}(2M^{2}-m_{b}^{2})(1+2%
\beta )  \notag \\
&+&6I_{1}\Big(\mathcal{V}_{\parallel }(\alpha ),1\Big)%
f_{K}m_{K}^{2}(2M^{2}-m_{b}^{2})(1+\beta )+8I_{1}\Big(\mathcal{V}_{\perp
}(\alpha ),1\Big)f_{K}m_{K}^{2}(2M^{2}-m_{b}^{2})(2+\beta )  \notag \\
&+&8I_{1}\Big(\mathcal{T}(\alpha ),1\Big)\mu _{K}m_{K}^{2}m_{b}(1+5\beta
)u_{0}-16I_{1}\Big(\mathcal{T}(\alpha ),v\Big)\mu
_{K}m_{K}^{2}m_{b}(1+2\beta )u_{0}  \notag \\
&-&12I_{1}\Big(\mathcal{A}_{\parallel }(\alpha ),v\Big)%
f_{K}m_{K}^{2}(2M^{2}-m_{b}^{2})(1+\beta )-16I_{1}\Big(\mathcal{V}_{\perp
}(\alpha ),v\Big)f_{K}m_{K}^{2}(2M^{2}-m_{b}^{2})(2+\beta )  \notag \\
&+&2I_{2}\Big(\mathcal{T}(\alpha ),1\Big)\mu _{K}M^{2}m_{b}(1+5\beta )-4I_{2}%
\Big(\mathcal{T}(\alpha ),v\Big)\mu _{K}M^{2}m_{b}(1+2\beta )+\dfrac{%
2f_{K}m_{K}^{2}m_{b}^{3}m_{s}}{M^{2}}\Bigg[\Big(\gamma _{E}(7\beta
-1)-6\beta \Big)  \notag \\
&\times &I_{1}\Big(\mathcal{A}_{\parallel }(\alpha ),1\Big)+\Big(6\gamma
_{E}(1+3\beta )-4(1+2\beta )\Big)I_{1}\Big(\mathcal{A}_{\perp }(\alpha ),1%
\Big)+2\Big(\gamma _{E}(5+\beta )-2(2+\beta )\Big)I_{1}\Big(\mathcal{V}%
_{\perp }(\alpha ),1\Big)  \notag \\
&+&3\Big(\gamma _{E}(3+\beta )-2\Big)I_{1}\Big(\mathcal{V}_{\parallel
}(\alpha ),1\Big)+6\beta (2-3\gamma _{E})I_{1}\Big(\mathcal{A}_{\parallel
}(\alpha ),v\Big)+4(2+\beta )(2-3\gamma _{E})_{1}\Big(\mathcal{V}_{\perp
}(\alpha ),v\Big)\Bigg]\Bigg]\Bigg\},  \label{eq:PiGG}
\end{eqnarray}%
\begin{eqnarray}
\widetilde{\Pi }^{\mathrm{\langle sG\bar{s}\rangle }} &=&\frac{%
m_{0}^{2}\langle s\bar{s}\rangle }{864\sqrt{2}M^{6}}e^{\frac{%
m_{K}^{2}-4m_{b}^{2}}{4M^{2}}}\Bigg\{-f_{K}m_{K}^{2}m_{b}^{2}\Big[9\beta
m_{b}^{2}+M^{2}(v(7+2\beta )+\beta -1)\Big]\mathbb{A}(u_{0})+4M^{2}\Big[%
f_{K}M^{2}\Big(9\beta m_{b}^{2}  \notag \\
&&+M^{2}(2\beta (5+v)+7v-1)\Big)\phi _{K}(u_{0})-3\mu _{K}(\widetilde{\mu }%
_{K}^{2}-1)(\beta -1)m_{b}^{3}\phi _{\sigma }(u_{0})\Big]+\dfrac{m_{b}m_{s}}{%
M^{2}}\Bigg[f_{K}m_{K}^{2}\Big((1-\beta )M^{4}  \notag \\
&&+3(1+\beta )m_{b}^{4}-(7+5\beta )M^{2}m_{b}^{2}\Big)\mathbb{A}%
(u_{0})-4f_{K}M^{4}\Big((\beta -1)M^{2}+3(1+\beta )m_{b}^{2}\Big)\phi
_{K}(u_{0})+2\mu _{K}(\tilde{\mu}_{K}^{2}-1)  \notag \\
&&M^{2}m_{b}\Big(2\beta (M^{2}+m_{b}^{2})-M^{2}\Big)\phi _{\sigma }(u_{0})%
\Bigg]+6M^{2}f_{K}m_{K}^{2}m_{b}^{2}\Big[3\beta I_{1}\Big(\mathcal{A}%
_{\parallel }(\alpha ),1\Big)+2(1+2\beta )I_{1}\Big(\mathcal{A}_{\perp
}(\alpha ),1\Big)  \notag \\
&&+3I_{1}\Big(\mathcal{V}_{\parallel }(\alpha ),1\Big)+2(2+\beta )I_{1}\Big(%
\mathcal{V}_{\perp }(\alpha ),1\Big)-6\beta I_{1}\Big(\mathcal{A}_{\parallel
}(\alpha ),v\Big)-4(2+\beta )I_{1}\Big(\mathcal{V}_{\perp }(\alpha ),v\Big)%
\Big]  \notag \\
&&+2f_{K}m_{K}^{2}m_{s}(1-\beta )(M^{2}-m_{b}^{2})\Big[I_{1}\Big(\mathcal{A}%
_{\parallel }(\alpha ),1\Big)+2I_{1}\Big(\mathcal{V}_{\perp }(\alpha ),1\Big)%
\Big]+16\mu _{K}m_{K}^{2}m_{b}m_{s}u_{0}  \notag \\
&&\times \Big[(1+2\beta )I_{1}\Big(\mathcal{T}(\alpha ),1\Big)-(2+\beta
)I_{1}\Big(\mathcal{T}(\alpha ),v\Big)\Big]+4\mu _{K}M^{2}m_{b}m_{s}\Big[%
(1+2\beta )I_{2}\Big(\mathcal{T}(\alpha ),1\Big)  \notag \\
&&-(2+\beta )I_{2}\Big(\mathcal{T}(\alpha ),v\Big)\Big]\Bigg\},
\label{eq:PiqGqbar}
\end{eqnarray}%
\begin{eqnarray}
\widetilde{\Pi }^{\mathrm{\langle s\bar{s}\rangle \langle GG\rangle }%
}(M^{2}) &=&\dfrac{\langle s\bar{s}\rangle \langle GG\rangle }{5184\sqrt{2}%
\pi ^{2}M^{8}}e^{\frac{m_{K}^{2}-4m_{b}^{2}}{4M^{2}}}m_{b}\Bigg\{3\beta
f_{K}m_{K}^{2}m_{b}(2M^{2}-m_{b}^{2})\mathbb{A}(u_{0})+4M^{2}\Big[3\beta
M^{2}f_{K}m_{b}\phi _{K}(u_{0})  \notag \\
&+&(\widetilde{\mu }_{K}^{2}-1)\mu _{K}(m_{b}^{2}-3M^{2})(\beta -1)\phi
_{\sigma }(u_{0})\Big]+\dfrac{m_{s}}{2M^{2}}\Bigg[(1+\beta )f_{K}\Big(%
m_{K}^{2}(m_{b}^{4}-6M^{2}m_{b}^{2}+6M^{4})\mathbb{A}(u_{0})  \notag \\
&-&4M^{4}(m_{b}^{2}-3M^{2})\phi _{K}(u_{0})\Big)+4\beta \mu _{K}(\tilde{\mu}%
_{K}^{2}-1)M^{2}m_{b}\phi _{\sigma }(u_{0})\Bigg]\Bigg\},
\label{eq:PiqqbarGG}
\end{eqnarray}%
\begin{eqnarray}
\widetilde{\Pi }^{\mathrm{\langle sG\bar{s}\rangle \langle GG\rangle }%
}(M^{2}) &=&\frac{m_{0}^{2}\langle s\bar{s}\rangle \langle GG\rangle }{20736%
\sqrt{2}M^{12}}e^{\frac{m_{K}^{2}-4m_{b}^{2}}{4M^{2}}}m_{b}\Bigg\{3\beta
f_{K}m_{K}^{2}m_{b}\Big[m_{b}^{2}(m_{b}^{2}-6M^{2})+6M^{4}\Big]\mathbb{A}%
(u_{0})-4M^{2}\Big[3\beta M^{2}f_{K}m_{b}  \notag \\
&&\times (m_{b}^{2}-2M^{2})\phi _{K}(u_{0})+(\widetilde{\mu }_{K}^{2}-1)\mu
_{K}(\beta -1)\Big(m_{b}^{2}(m_{b}^{2}-6M^{2})+6M^{4}\Big)\phi _{\sigma
}(u_{0})\Big]  \notag \\
&&+\dfrac{m_{s}}{3M^{2}}\Bigg[3(1+\beta
)f_{K}m_{K}^{2}(3M^{2}-m_{b}^{2})(m_{b}^{4}-8M^{2}m_{b}^{2}+6M^{4})\mathbb{A}%
(u_{0})+4M^{2}(m_{b}^{4}-6M^{2}m_{b}^{2}+6M^{4})  \notag \\
&&\times \Big(3(1+\beta )f_{K}M^{2}\phi _{K}(u_{0})-\beta \mu _{K}(\tilde{\mu%
}_{K}^{2}-1)m_{b}\phi _{\sigma }(u_{0})\Big)\Bigg]\Bigg\}.
\label{eq:PiqGqbarGG}
\end{eqnarray}

In Eqs.\ (\ref{PiQCD})-(\ref{eq:PiqGqbarGG}) the following shorthand
notations are used:
\begin{eqnarray}
I_{1}\Big(\Phi (\alpha ),f(v)\Big) &=&\int \mathcal{D}\alpha
_{i}\int_{0}^{1}dv\Phi (\alpha _{\bar{q}},\alpha _{q},\alpha _{g})f(v)\delta
(k-u_{0}),  \notag \\
I_{2}\Big(\Phi (\alpha ),f(v)\Big) &=&\int \mathcal{D}\alpha
_{i}\int_{0}^{1}dv\Phi (\alpha _{\bar{q}},\alpha _{q},\alpha _{g})f(v)\delta
^{^{\prime }}(k-u_{0}),  \label{eq:I1}
\end{eqnarray}%
and
\begin{equation*}
\Psi =\dfrac{M^{2}(s-m_{b}^{2})}{s\Lambda ^{2}},\ \mu _{K}=\frac{%
f_{K}m_{K}^{2}}{m_{s}+m_{u}},~~~~~\widetilde{\mu }_{K}=\frac{m_{s}+m_{u}}{%
m_{K}},~k=\alpha _{q}+\alpha _{g}v.
\end{equation*}%
In expressions above $u_{0}=1/2,$ $\gamma _{E}=0.557721$ is the
Euler-Mascheroni constant, and $\Lambda $ is the QCD scale parameter.

Equations (\ref{PiQCD})-(\ref{eq:PiqGqbarGG}) depend on various DAs of $K$
meson. We take into account two- and three-particle distributions up to
twist-4. The DAs which appear in the equalities above are given by the
following expressions \cite{Rde09}:
\begin{eqnarray}
\phi _{\mathrm{K}}(u) &=&6u\bar{u}\left[ 1+a_{1}^{\mathrm{K}%
}C_{1}^{3/2}(2u-1)+a_{2}^{\mathrm{K}}C_{2}^{3/2}(2u-1)\right] ,  \notag
\label{ede31} \\
\mathcal{T}(\alpha _{i}) &=&360\eta _{3}\alpha _{\bar{q}}\alpha _{q}\alpha
_{g}^{2}\left[ 1+w_{3}{\frac{1}{2}}(7\alpha _{g}-3)\right] ,  \notag \\
\phi _{\sigma }(u) &=&6u\bar{u}\left[ 1+\left( 5\eta _{3}-{\frac{1}{2}}\eta
_{3}w_{3}-{\frac{7}{20}}\mu _{K}^{2}-{\frac{3}{5}}\mu _{K}^{2}a_{2}^{\mathrm{%
K}}\right) C_{2}^{3/2}(2u-1)\right] ~,  \notag \\
\mathcal{V}_{\parallel }(\alpha _{i}) &=&120\alpha _{q}\alpha _{\bar{q}%
}\alpha _{g}\left[ v_{00}+v_{10}(3\alpha _{g}-1)\right] ,\ \ \ \mathcal{A}%
_{\parallel }(\alpha _{i})=120\alpha _{q}\alpha _{\bar{q}}\alpha _{g}\left[
0+a_{10}(\alpha _{q}-\alpha _{\bar{q}})\right] ,  \notag \\
\mathcal{V}_{\perp }(\alpha _{i}) &=&-30\alpha _{g}^{2}\left\{
h_{00}(1-\alpha _{g})+h_{01}\left[ \alpha _{g}(1-\alpha _{g})-6\alpha
_{q}\alpha _{\bar{q}}\right] +h_{10}\left[ \alpha _{g}(1-\alpha _{g})-{\frac{%
3}{2}}(\alpha _{\bar{q}}^{2}+\alpha _{q}^{2}\right] \right\} ,  \notag \\
\mathcal{A}_{\perp }(\alpha _{i}) &=&30\alpha _{g}^{2}(\alpha _{\bar{q}%
}-\alpha _{q})\left[ h_{00}+h_{01}\alpha _{g}+{\frac{1}{2}}h_{10}(5\alpha
_{g}-3)\right] ~,  \notag \\
\mathbb{A}(u) &=&6u\bar{u}\left[ {\frac{16}{15}}+{\frac{24}{35}}a_{2}^{%
\mathrm{K}}+20\eta _{3}+{\frac{20}{9}}\eta _{4}+\left( -{\frac{1}{15}}+{%
\frac{1}{16}}-{\frac{7}{27}}\eta _{3}w_{3}-{\frac{10}{27}}\eta _{4}\right)
C_{2}^{3/2}(2u-1)\right.  \notag \\
&&\left. +\left( -{\frac{11}{210}}a_{2}^{\mathrm{K}}-{\frac{4}{135}}\eta
_{3}w_{3}\right) C_{4}^{3/2}(2u-1)\right] ~,  \notag \\
&+&\left( -{\frac{18}{5}}a_{2}^{\mathrm{K}}+21\eta _{4}w_{4}\right) \left[
2u^{3}(10-15u+6u^{2})\ln u\right.  \notag \\
&&\left. +2\bar{u}^{3}(10-15\bar{u}+6\bar{u}^{2})\ln \bar{u}+u\bar{u}(2+13u%
\bar{u})\right] ~,
\end{eqnarray}%
where $C_{n}^{k}(x)$ are the Gegenbauer polynomials, and
\begin{eqnarray}
h_{00} &=&v_{00}=-{\frac{1}{3}}\eta _{4},~\ a_{10}={\frac{21}{8}}\eta
_{4}w_{4}-{\frac{9}{20}}a_{2}^{\mathrm{K}},\ \ v_{10}={\frac{21}{8}}\eta
_{4}w_{4},\ \ h_{01}={\frac{7}{4}}\eta _{4}w_{4}-{\frac{3}{20}}a_{2}^{%
\mathrm{K}}~,  \notag \\
h_{10} &=&{\frac{7}{4}}\eta _{4}w_{4}+{\frac{3}{20}}a_{2}^{\mathrm{K}}~,\
g_{0}=1,\ g_{2}=1+{\frac{18}{7}}a_{2}^{\mathrm{K}}+60\eta _{3}+{\frac{20}{3}}%
\eta _{4},\ g_{4}=-{\frac{9}{28}}a_{2}^{\mathrm{K}}-6\eta _{3}w_{3}.~
\label{eq:ede32}
\end{eqnarray}

\end{widetext}

The parameters $a_{1}^{\mathrm{K}}=0.06\pm 0.03$ and $a_{2}^{\mathrm{K}%
}=0.25\pm 0.15$ are borrowed from Ref.\ \cite{Ball:2006wn}, whereas for
decay constant of $K$ meson $f_{K}=0.16~\mathrm{GeV}$, and for $\eta
_{3}=0.015$, $\eta _{4}=0.6$, $w_{3}=-3$, $w_{4}=0.2$ we use estimations
from Ref.\ \cite{Rde09}. Information on other distribution amplitudes of $K$
meson can be found in Refs.\ \cite{Rde09,Ball:2006wn}.

Here we have also collected formulas, which can be applied in the continuum
subtraction. In the left-hand side of the formulas we present the original
forms as they appear after double Borel transformation, whereas in the
right-hand side we provide their subtracted version used in sum rule
calculations:
\begin{equation}
\left( M^{2}\right) ^{N}\int_{m^{2}}^{\infty }dse^{-s/M^{2}}f(s)\rightarrow
\int_{m^{2}}^{s_{0}}dse^{-s/M^{2}}F_{N}(s).
\end{equation}%
For the more complicated case
\begin{equation}
\left( M^{2}\right) ^{N}\ln \left( \frac{M^{2}}{\Lambda ^{2}}\right)
\int_{m^{2}}^{\infty }dse^{-s/M^{2}}f(s),
\end{equation}%
for all values of $N$ the following expression is applicable
\begin{eqnarray}
&& \int_{m^{2}}^{s_{0}}dse^{-s/M^{2}}\left[ F_{N}(m^{2})\ln \left( \frac{%
s-m^{2}}{\Lambda ^{2}}\right) +\gamma _{E}F_{N}(s) \right.  \notag \\
&&\left. +\int_{m^{2}}^{s}duF_{N-1}(u)\ln \left( \frac{s-u}{\Lambda ^{2}}%
\right) \right] .
\end{eqnarray}

The next formula is
\begin{eqnarray}
&&\left( M^{2}\right) ^{N}\ln \left( \frac{M^{2}}{\Lambda ^{2}}\right)
e^{-m^{2}/M^{2}}  \notag \\
&&\rightarrow e^{-s_{0}/M^{2}}\sum_{i=1}^{1-N}\left( \frac{d}{ds_{0}}\right)
^{1-N-i}\left[ \ln \left( \frac{s_{0}-m^{2}}{\Lambda ^{2}}\right) \right]
\frac{1}{\left( M^{2}\right) ^{i-1}}  \notag \\
&&+\gamma _{E}\left( M^{2}\right) ^{N}\left(
e^{-m^{2}/M^{2}}-\delta_{N1}e^{-s_{0}/M^{2}}\right)  \notag \\
&&+\left( M^{2}\right)^{N-1}\int_{m^{2}}^{s_{0}}dse^{-s/M^{2}}\ln \left(
\frac{s-m^{2}}{\Lambda ^{2}}\right),
\end{eqnarray}%
if $N\leq 1$, and
\begin{eqnarray}
&&\frac{\gamma _{E}}{\Gamma (N)}\int_{m^{2}}^{s_{0}}dse^{-s/M^{2}}\left(
s-m^{2}\right) ^{N-1}  \notag \\
&&+\frac{1}{\Gamma (N-1)}\int_{m^{2}}^{s_{0}}dse^{-s/M^{2}}%
\int_{m^{2}}^{s}du(s-u)^{N-2}  \notag \\
&&\times \ln \left( \frac{u-m^{2}}{\Lambda ^{2}}\right),
\end{eqnarray}%
for $N>1$.

It is worth to note also the expressions
\begin{eqnarray}
&&\left( M^{2}\right) ^{N}\int_{m^{2}}^{\infty }dse^{-s/M^{2}}f(s)\ln \left(
\frac{s-m^{2}}{\Lambda ^{2}}\right)  \notag \\
&&\rightarrow e^{-s_{0}/M^{2}}\sum_{i=1}^{|N|}\frac{\widetilde{F}%
_{N+i}(s_{0})}{\left( M^{2}\right) ^{i-1}}+\left( M^{2}\right)
^{N}\int_{m^{2}}^{s_{0}}dse^{-s/M^{2}}f(s)  \notag \\
&&\times \ln \left( \frac{s-m^{2}}{\Lambda ^{2}}\right) ,\ \ \ N\leq 0,
\end{eqnarray}%
and%
\begin{eqnarray}
&&\frac{1}{\Gamma (N)}\int_{m^{2}}^{s_{0}}dse^{-s/M^{2}}%
\int_{m^{2}}^{s}du(s-u)^{N-1}  \notag \\
&&\times \ln \left( \frac{u-m^{2}}{\Lambda ^{2}}\right)f(u) ,\ \ N>0.
\end{eqnarray}%
In the equations above we have employed the notations%
\begin{equation}
F_{N}(s)=\left( \frac{d}{ds}\right)^{-N}f(s), \,\,\, N\leq 0,
\end{equation}%
and%
\begin{equation}
F_{N}(s)=\frac{1}{\Gamma (N)}\int_{m^{2}}^{s}du(s-u)^{N-1}f(u),\, \,\,N>0.
\end{equation}%
For $N\leq 0$ we have also used:
\begin{eqnarray}
&&\widetilde{F}_{N}(s)=\left( \frac{d}{ds}\right) ^{-N}\left[
f(s)\int_{1}^{\infty }\frac{dt}{t}\exp \left( -\frac{\Lambda ^{2}t}{s-m^{2}}%
\right) \right],  \notag \\
&&\widetilde{F}_{N}(s_{0})=\left( \frac{d}{ds_{0}}\right) ^{-N}\left[
f(s_{0})\ln \left( \frac{s_{0}-m^{2}}{\Lambda ^{2}}\right) -\gamma _{E}%
\right].  \notag \\
&&{}
\end{eqnarray}%
The expressions provided above are valid only if $f(m^2)=0$. In other cases,
one has to use the prescription $f(s)=[f(s)-f(m^2)]+f(m^2)$, where the first
term in the brackets is equal to zero, when $s=m^2$.

\end{document}